\documentclass[10pt,journal,compsoc]{IEEEtran}

\ifCLASSOPTIONcompsoc
  \usepackage[nocompress]{cite}
\else
  \usepackage{cite}
\fi
\ifCLASSINFOpdf
\else
\fi

\usepackage{times}
\usepackage{epsfig}
\usepackage{graphicx}
\usepackage{comment}
\usepackage{amsmath,amssymb} 
\usepackage{color}
\usepackage{booktabs}
\usepackage{multirow}
\usepackage[utf8]{inputenc}
\usepackage{pifont}
\usepackage{microtype}
\usepackage{subcaption}
\usepackage{tabularx}
\usepackage{xcolor,soul}

\definecolor{blue}{rgb}{1,1,1}
\sethlcolor{blue}

\definecolor{cornflower}{rgb}{0.5,0.73,0.90}

\DeclareRobustCommand{\hlred}[1]{{\sethlcolor{white}\hl{#1}}}

\newcommand{\sq}[1]{\textcolor{black}{#1}}

\soulregister\cite7 
\soulregister\citep7
\soulregister\citet7
\soulregister\ref7 
\soulregister\pageref7 

\begin{document}
%
\title{Meta-PU: An Arbitrary-Scale Upsampling Network for Point Cloud}

\author{Shuquan~Ye, Dongdong~Chen, Songfang~Han, Ziyu~Wan and Jing~Liao
\IEEEcompsocitemizethanks{\IEEEcompsocthanksitem Shuquan Ye, Ziyu Wan and Jing Liao are with the Department of Computer Science, City University of Hong Kong, HK, China.\protect\\
E-mails: shuquanye2-c@my.cityu.edu.hk, ziyuwan2-c@my.cityu.edu.hk and jingliao@cityu.edu.hk
\IEEEcompsocthanksitem Dongdong Chen is with Microsoft Research. \protect\\E-mail: cddlyf@gmail.com
\IEEEcompsocthanksitem Songfang Han is with University of California, San Diego, USA. \protect\\E-mail: hansongfang@gmail.com
\IEEEcompsocthanksitem Jing Liao is the corresponding author.}
}

%
%

\markboth{Journal of \LaTeX\ Class Files,~Vol.~14, No.~8, August~2015}%
{Shell \MakeLowercase{\textit{et al.}}: Bare Advanced Demo of IEEEtran.cls for IEEE Computer Society Journals}
%



\IEEEtitleabstractindextext{%
\begin{abstract}
Point cloud upsampling is vital for the quality of \hlred{the} mesh in \hlred{three-dimensional} reconstruction. Recent research on point cloud upsampling has achieved great success \hlred{due} to the development of deep learning. However, \hlred{the} existing methods regard point cloud upsampling of different scale factors as independent tasks\hlred{. T}hus\hlred{, the methods} need to train a specific model for each scale factor, which is both inefficient and impractical \hlred{for storage and computation} in real applications. To address this limitation, in this work, we propose a novel method called ``Meta-PU" to firstly support point cloud upsampling of arbitrary scale factors with a single model. In \hlred{the} Meta-PU \hlred{method}, besides the backbone network consisting of residual graph convolution (RGC) blocks, a \hlred{meta-subnetwork} is learned to adjust the weights of \hlred{the} RGC block\hlred{s} dynamically, and a farthest sampling block is adopted to sample different numbers of points. \hlred{Together, these two blocks} enable our Meta-PU to continuously upsample \hlred{the} point cloud with arbitrary scale factors by \hlred{using only} a single model. In addition, \hlred{the} experiments \hlred{reveal} that training \hlred{on} multiple scales simultaneously is beneficial to each other. Thus\hlred{, }Meta-PU even outperforms \hlred{the} existing methods\hlred{ }trained for a specific scale factor only. 
\end{abstract}

}

\maketitle

\IEEEdisplaynontitleabstractindextext

%
\IEEEpeerreviewmaketitle

\ifCLASSOPTIONcompsoc
\IEEEraisesectionheading{\section{Introduction}\label{sec:introduction}}
\else
\section{Introduction}
\label{sec:introduction}
\fi

\IEEEPARstart{P}{oint} clouds are the most fundamental and popular representation for \hlred{three-dimensional (3D)} environment modeling. When reconstructing the 3D model of an object from the real world, a common \hlred{technique} is\hlred{ }to obtain the point cloud and then recover the mesh from it. However, a raw point cloud generated from depth cameras or reconstruction algorithms is usually sparse and noisy due to the restrictions of hardware devices or the limitations of algorithms, which leads to a low-quality mesh. To solve this problem, it is common\hlred{ }to apply point cloud upsampling prior to meshing, which takes a set of sparse points as input and generates a denser set of points to reflect the underlying surface better.

Conventional point cloud upsampling methods \cite{1175093,Huang2009,Wu:2015:DPC:2816795.2818073} are optimization-based with various shape priors as constraints, such as the local smoothness of the surface and \hlred{the} normal. These works perform\hlred{ }well for simple objects\hlred{ }but are not able to handle complex and dedicated structures. \hlred{Due} to the success of deep learning, some data-driven methods \cite{yu2018pu,Yifan_2019_CVPR,Li_2019_ICCV,wu2019point} \hlred{have emerged} recently and achieve\hlred{d} state-of-the-art performance by employing powerful deep neural networks to learn the upsampling process in an end-to-end way.

\begin{figure}[ht] 
\begin{center}
\includegraphics[width=0.45\textwidth]{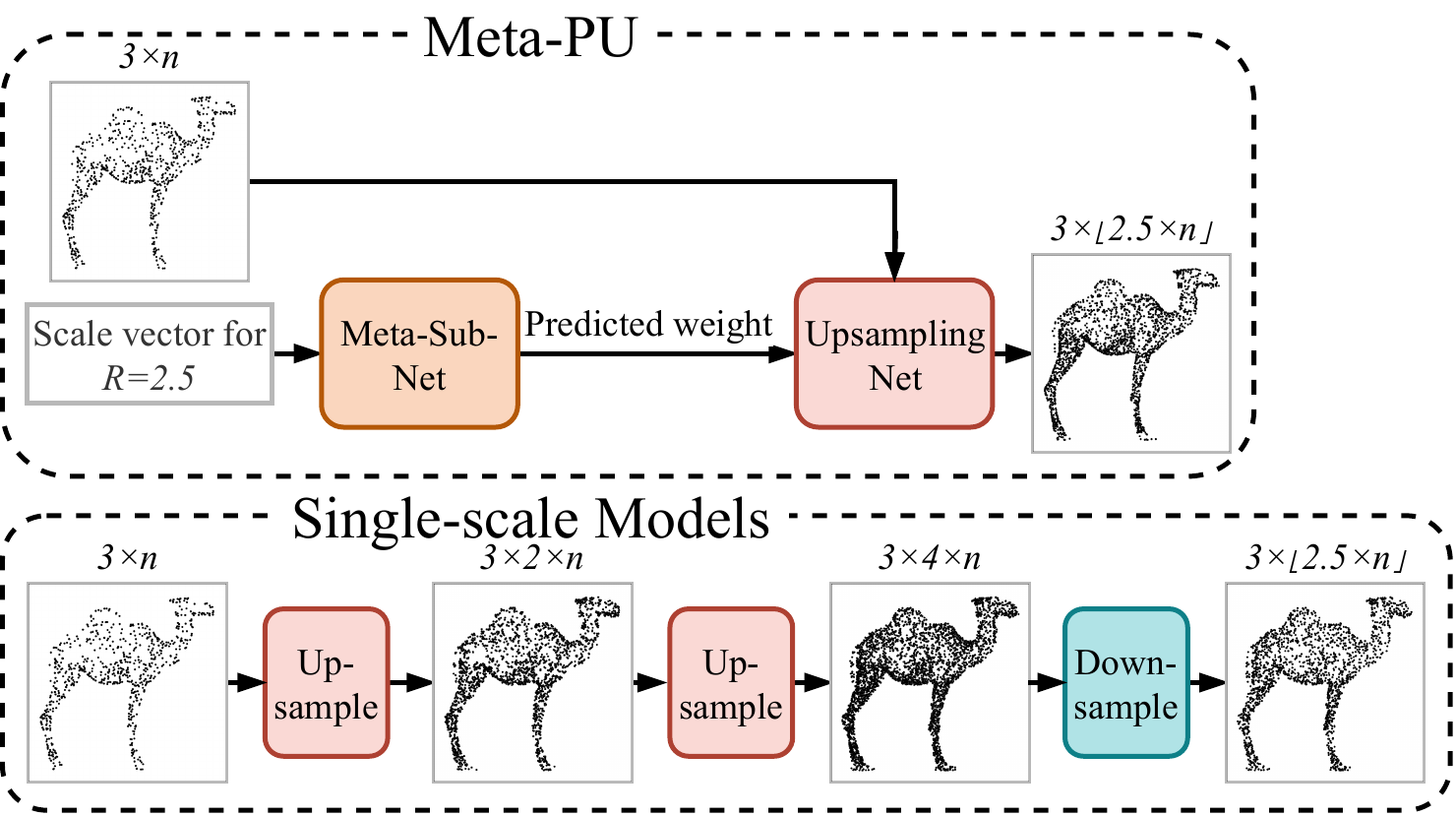}
\end{center}

\caption{\hlred{Arbitrary-scale} model Meta-PU vs\hlred{. the} single-scale models over \hlred{the} example scale $R=2.5$. \hlred{The existing single-scale models} first need to scale to a larger integer scale (e.g., 4x), then use a downsample algorithm to achieve the \hlred{noninteger} scale \hlred{of} 2.5x.}
\label{teaser}
\end{figure}

However, all existing point cloud upsampling networks only consider certain integer scale factors (e.g., 2x). They regard the upsampling of different scale factors as independent tasks. Thus\hlred{,} a specific model for each scale factor has to be trained, \hlred{limiting} the use of these methods in real-world scenarios where different scale factors are needed to fit different densities of\hlred{ }raw point clouds. Some works \cite{Li_2019_ICCV,wu2019point} suggests to achieve \hlred{larger} scales \hlred{through} iterative upsampling\hlred{ (}e.g., upsampling 4x by running the 2x model twice\hlred{)}. However, this repeated computation is time-consuming, and upsampling of non-integer factors still cannot be achieved\hlred{ (}e.g., 2.5x with single-scale models\hlred{)}, as \hlred{depicted} in Fig.\hlred{ }\ref{teaser}.

In real-world scenarios, it is very common and necessary to upsample\hlred{ }raw point clouds into various user-customized densities for mesh reconstruction, point cloud processing, or other needs. Thus\hlred{,} an efficient method for\hlred{ }upsampling\hlred{ }arbitrary scale factors is desired to solve the aforementioned drawbacks in \hlred{the} existing methods. However, it is not easy for vanilla neural networks. Their behavior is fixed once trained because of the deterministic learned weights, so it is not\hlred{ }straightforward to let the network handle the arbitrary scale factor on the fly.

Motivated by the development of meta-learning \cite{fan2018decouple,fan2019general} and the latest image super-resolution method \cite{hu2019meta}, we propose an efficient and novel network called ``Meta-PU" for point upsampling of arbitrary scale factors. By incorporating one extra cheap \hlred{meta-subnetwork} as the controller,\hlred{ }Meta-PU can dynamically change its behavior during runtime depending on the desired scale factor. Compared with storing the weights for individual scale factors, storing the \hlred{meta-subnetwork} is more convenient and flexible.

Specifically, the backbone of\hlred{ }Meta-PU is based on a graph convolution\hlred{al} network (GCN), consisting of several residual graph convolution\hlred{al} (RGC) blocks to extract \hlred{the} feature representation of each point as well as relationships to their nearest neighbors. And the \hlred{meta-subnetwork} is trained to generate weights for the \hlred{meta-RGC block} given \hlred{the} input of a scale factor. Then\hlred{,} the \hlred{m}eta-convolution uses these weights to extract features that are \hlred{adaptively} tailored \hlred{to the} scale factor. Following several RGC blocks, a farthest sampling block is further added to output \hlred{an} arbitrary number of points. In this way, different scale factors can be trained simultaneously with a single model. At the inference stage, when users specify an upsampling scale factor, the \hlred{meta-subnetwork} will dynamically change the behavior of the \hlred{meta-RGC block} by adapting its weights and output\hlred{s} the corresponding upsampling results.

To demonstrate the effectiveness and flexibility of our method, we compare it with several strong baseline methods. \hlred{The comparison} shows that our method can even achieve SOTA performances for specific single scale factors while supporting arbitrary\hlred{-}scale upsampling for the first time. In other words, our approach is both stronger and more flexible than the SOTA approaches. \hlred{To }better \hlred{understand} the underlying working principle and broader applications, we further provide a comprehensive analysis from different perspectives. 

In summary, our contribution \hlred{is three-fold}:
\begin{itemize}
\item We propose the first point cloud upsampling network \hlred{that} supports arbitrary scale factors (including \hlred{noninteger} factors), via a meta-learning approach.

\item We show that jointly training multiple scale factors with one model \hlred{improves} performance. Our arbitrary-scale model even achieves better results at each specific scale than the single-scale counterpart.

\item We evaluate our method on multiple benchmark datasets and demonstrate that\hlred{ }Meta-PU advances\hlred{ }state-of-the-art performance.
\end{itemize}

\section{Related Work}

\noindent\textbf{Optimization-based upsampling.} Point cloud upsampling is formulated as an optimization problem in early \hlred{work.} A pioneering solution proposed by Alexa \emph{et al.}~\cite{1175093} constructs \hlred{a} Voronoi diagram on the surface and then inserts points at \hlred{the} vertices. Lipman \emph{et al.}~\cite{lipman2007parameterization} design\hlred{ed} a novel locally optimal projection\hlred{ }operator for \hlred{point} resampling and surface reconstruction based on $L_1$ median, which is robust to noise outliers. Later Huang \emph{et al.}~\cite{Huang2009} improve\hlred{d the locally optimal projection} operator to enable\hlred{ }edge-aware point set upsampling. Wu \emph{et al.}~\cite{Wu:2015:DPC:2816795.2818073} employ\hlred{ed} a joint optimization method for \hlred{the} inner points and surface points defined in their new point set representation. However, most of these methods have a strong a priori assumption\hlred{ (}e.g., a reasonable normal estimation or a smooth surface in the local geometry\hlred{).} Thus, they may easily suffer from complex and massive point cloud data.

\vspace{1em}
\noindent\textbf{Deep-learning-based upsampling.} Recently, deep learning has become a powerful tool for extracting features directly from point cloud data in a data-driven way. Qi \emph{et al.} firstly propose PointNet~\cite{Qi_2017_CVPR} and PointNet++~\cite{NIPS2017_7095} for extracting multi-level features from point sets. Based on these flexible feature extractors, deep neural networks have been applied to many point cloud tasks, such as \hlred{those in} \cite{8805456,8388294,8730533}. As for\hlred{ }point cloud upsampling, Yu \emph{et al.}~\cite{yu2018pu} present\hlred{ed} a point cloud upsampling neural network operating on \hlred{the} patch level and \hlred{made} it possible to directly input \hlred{a }high-resolution point cloud. Then\hlred{,} Yu \emph{et al.} develop\hlred{ed} EC-Net~\cite{Yu_2018_ECCV} to improve the quality of the upsampled point cloud \hlred{using} an edge-aware joint learning strategy. Wang \emph{et al.} propose\hlred{d} a progressive point set upsampling network~\cite{Yifan_2019_CVPR} to\hlred{ }suppress noise \hlred{further} and preserve \hlred{the} details of the upsampled point cloud.
\hlred{Moreover}, different frameworks\hlred{, such as the} generative adversarial network (GAN) \cite{NIPS2014_5423} and \hlred{the} graph convolution\hlred{al} network(GCN)~\cite{kipf2017semi}\hlred{, have} attract\hlred{ed} researchers' attention for handling point cloud upsampling. Li \emph{et al.} propose\hlred{d the} PU-GAN~\cite{Li_2019_ICCV} \hlred{by} formulating a GAN framework to obtain more uniformly distributed point cloud results. Wu \emph{et al.} propose\hlred{d} AR-GCN~\cite{wu2019point} to make the first attempt to model point cloud upsampling into a \hlred{GCN}. However, these networks are only designed for\hlred{ }upsampling\hlred{ }a fixed scale factor. When different upsampling scales are required in practical applications, multiple models have to be retrained. \hlred{Unlike} their\hlred{ methods},\hlred{ }Meta-PU supports upsampling point clouds for arbitrary scale factors, by employing meta-learning to predict weights of the network and dynamically change\hlred{ }behavior for each scale factor.

\vspace{1em}
\noindent\textbf{Meta-learning.} \hlred{Meta-learning}, or learning to learn, refers to learning by observing the performance of different machine learning methods on various learning tasks. It is normally a two-level model\hlred{:} a meta-level model \hlred{performed} across tasks, and a base-level model acting within each task. The early meta-learning approach is \hlred{primarily} used in few-shot/zero-shot learning, and transfer learning \cite{andrychowicz2016learning,ravi2016optimization}. Recent works have also applied meta-learning to various tasks and achieved state-of-the-art results in object detection \cite{yang2018metaanchor}, instance segmentation \cite{hu2018learning}, image super-resolution \cite{hu2019meta}, image smoothing \cite{fan2018decouple,fan2019general,chen2020controllable}, network pruning\cite{liu2019metapruning}, etc. A more comprehensive survey of meta-learning can be found in \cite{lemke2015metalearning}. Among these works, \hlred{meta-SR} \cite{hu2019meta}\hlred{,} which learns the weights of the network for arbitrary-scale image super-resolution, is \hlred{closely} related to ours. However, it cannot be applied to our task. The main reason is that the target of \hlred{meta-SR} is images with \hlred{a} regular grid structure, \hlred{whereas} our\hlred{ target includes much more challenging} irregular and orderless point clouds. For regular grid-based images, since the correspondence between each output pixel and the corresponding input pixel is pre-determined, \hlred{meta-SR} can directly use\hlred{ }relative offset information to regress the local upsampling weights. \hlred{However, no such correspondence exists} for point clouds\hlred{.} Therefore, we resort to using the \hlred{meta-subnetwork} together with \hlred{the} farthest sampling block. The \hlred{meta-subnetwork} is responsible for adaptively tailoring the point features of a specific scale factor by dynamically adjusting the weights of the RGC block, while the sampling block is responsible for sampling\hlred{ }a particular number of points.

\begin{figure*}[ht] 
\begin{center}
\includegraphics[width=0.9\textwidth]{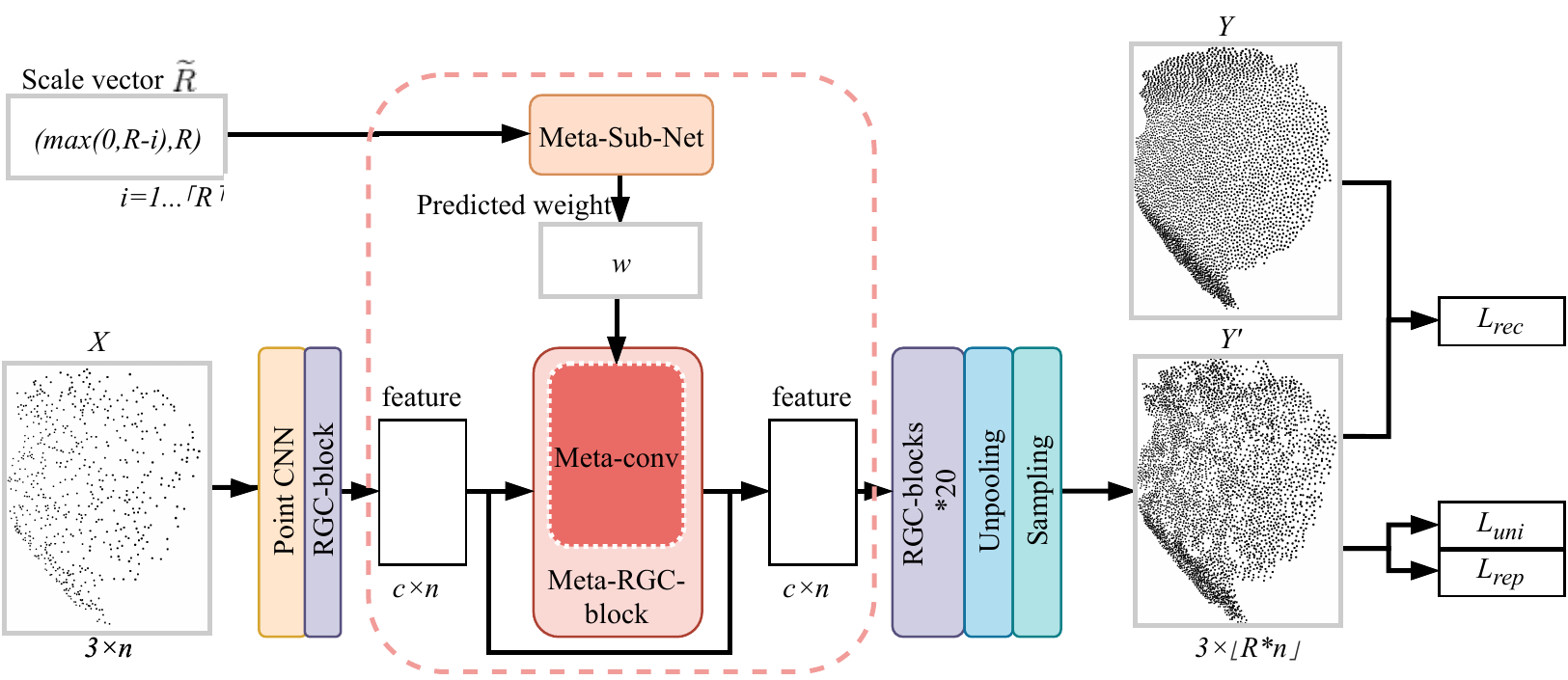}
\end{center}

  \caption{Overview of Meta-PU. Given a sparse input point cloud $X$ of $n$ points and \hlred{a} scale factor $R$, Meta-PU generates denser point cloud $Y'$ with $\left \lfloor R\times n \right \rfloor$ points. A compound loss function is employed to encourage $Y'$ to lie uniformly on the underlying surface of\hlred{ }target $Y$. The pink box is the core part of\hlred{ }Meta-PU. The \hlred{meta-subnetwork} takes scale factor $R$ as input and outputs \hlred{the} weight tensor $\mathbf{v}_{weight}$ for convolutional kernels in the meta-RGC block,\hlred{ }to \hlred{adapt }the feature extraction\hlred{ }to different upscales.}
\label{structure}
\end{figure*}

\section{Method}

In this section, we\hlred{ }define the task of arbitrary-scale point cloud upsampling\hlred{. T}hen we introduce the proposed Meta-PU in detail. 

\subsection{Arbitrary-scale Point Cloud Upsampling}

Given a sparse and unordered point set $ X =\left\{p_{i}\right\}_{i=1}^{n}$ of $n$ points, and \hlred{with a} scale factor\hlred{ }$R$, the task of arbitrary-scale point cloud upsampling is to generate a dense point set $ Y=\left\{p_{i}\right\}_{i=1}^{N}$ of $N=\left \lfloor R\times n \right \rfloor$ points. It is worth noting that $R$ is not necessarily an integer, and theoretically, $N$ can be any positive integer greater than $n$. The output $Y$ does not necessarily include\hlred{ }points in $X$. In addition, $X$ may not be uniformly distributed. In order to meet the needs of practical applications, we need the upsampled point cloud to satisfy the following two constraints. Firstly, each point of $Y$ lies on the underlying geometry surface described by $X$. Secondly, the distribution of output points should be smooth and uniform, \hlred{for any} scale factor $R$ or input point number $n$\hlred{.}

\subsection{Meta-PU}

\noindent\textbf{Overview.}  The backbone upsampling network contains five basic modules distinguished by different colors in Fig. \ref{structure}. The input point cloud first goes through a point \hlred{convolutional neural network (CNN)} and several \hlred{RGC} blocks to extract features for each centroid and its neighbors. Among these RGC blocks,\hlred{ the meta-RGC block} is special\hlred{. The meta-RGC block} weights are dynamically generated by a meta-sub-network given the input of $R$\hlred{. Thus }the features extracted by this meta-RGC block are tailored \hlred{to} the given scale factor. After \hlred{the} RGC blocks, an unpooling layer is followed to output $\left \lfloor R_{max}\times n \right \rfloor$ points, where $R_{max}$ denotes the maximum scale factor supported by our network, \hlred{and} $R_{max}=16$ by default. \hlred{Afterward}, the farthest sampling block is adopted to sample $N$ points from $\left \lfloor R_{max}\times n \right \rfloor$ points as the final output, which is constrained by a compound loss function. In the following section, we elaborate on the detailed structure of each block in Meta-PU, \hlred{and} the training loss.

\vspace{1em}
\noindent\textbf{Point CNN.} \hlred{The point CNN} is a simple structure on the spatial domain to extract features from the input point cloud $X$. In detail, for each point $p\in X$ with shape $1\times3$, we first group its $k$ nearest neighbors with shape $k\times 3$, \hl{and then feed them into a series of point-wise convolutions ($k\times c$) followed by a max-pooling layer to obtain $1\times c$ features, where $c$ is the channel number of point cloud feature}. \hlred{Thus,} the output feature $F_{out}$ is a tensor of shape $n \times c$. Recursively applied convolution reaches a wider receptive field representing more information, \hlred{whereas} the maximum pooling layer aggregates information from all points in the previous layer. In our implementation, we set $k=8$, and $c=128$ and the number of convolution\hlred{al} layers is $3$.


\begin{figure}[t]
\centering
\begin{subfigure}[RGC block]{0.48\textwidth}
\includegraphics[width=\textwidth]{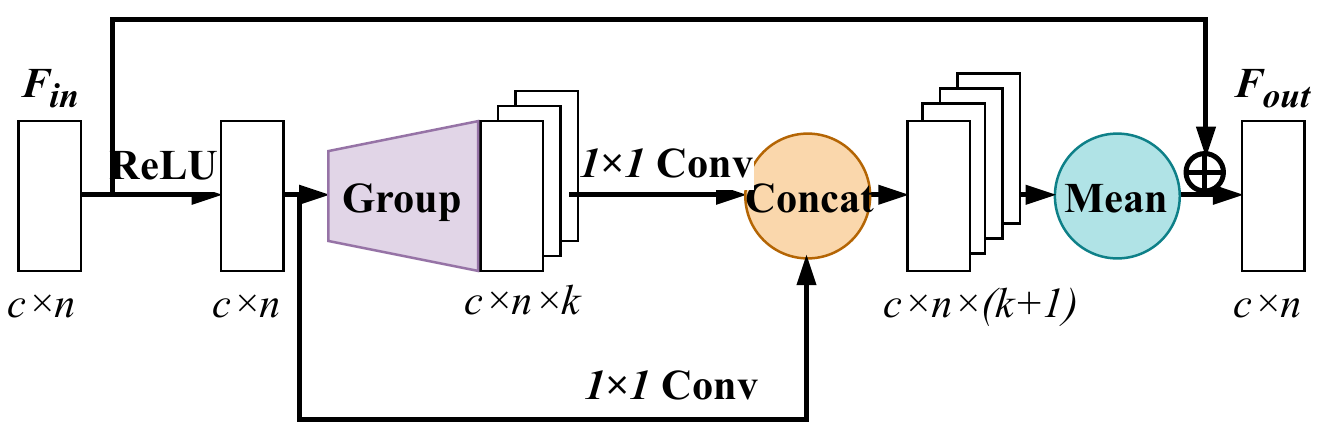}
\caption{RGC block}
\label{rgc}
\end{subfigure}
\begin{subfigure}[Meta-RGC block]{0.45\textwidth}
\includegraphics[width=\textwidth]{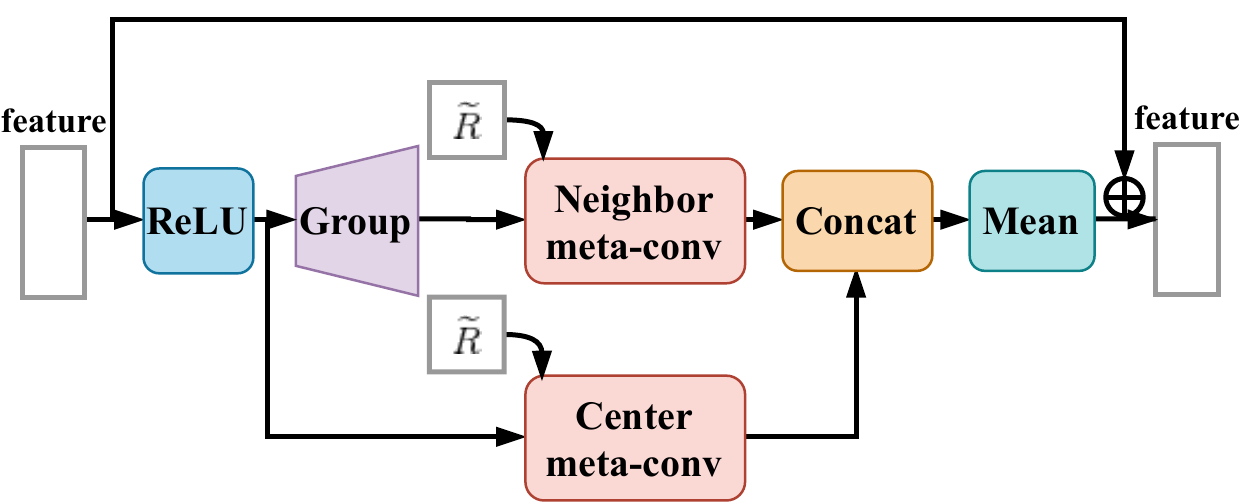}
\caption{Meta-RGC block}
\label{metargc}
\end{subfigure}
\begin{subfigure}[Unpooling block]{0.48\textwidth}
\includegraphics[width=\textwidth]{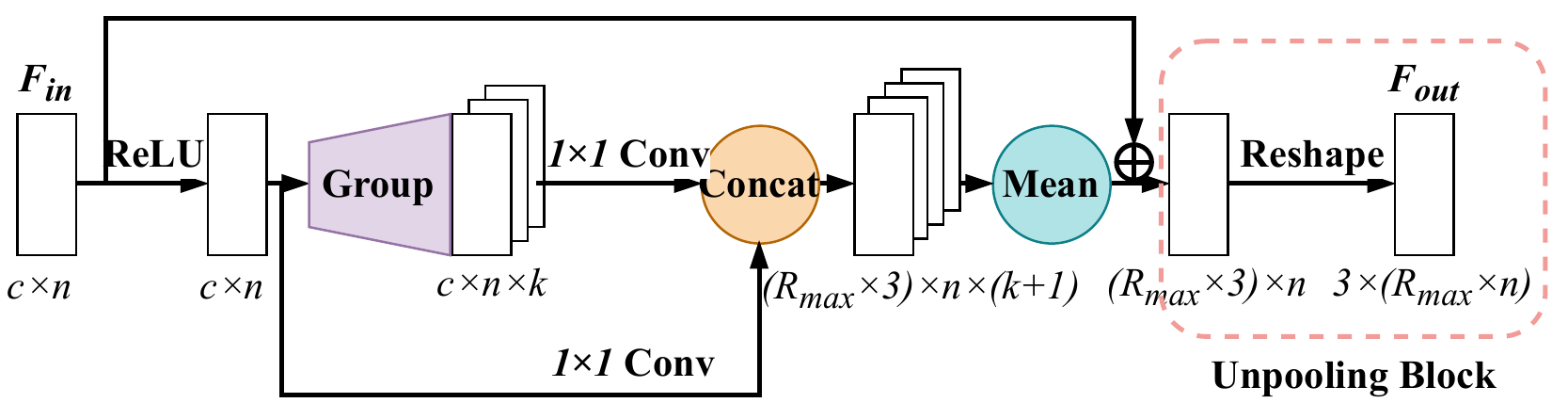}
\caption{Unpooling block}
\label{unpooling}
\end{subfigure}

\caption{Structure of RGC block (a), meta-RGC block (b) \hl{and unpooling block (c)}. Both RGC and meta-RGC blocks convolve centroid features and neighborhood features, respectively. \hlred{In} the meta-RGC block, the weights of its convolution\hlred{al} layers are dynamically predicted based on the scale factor $R$. \hl{The unpooling block follows the last RGC block.}}
\label{fig}

\end{figure}


\vspace{1em}
\noindent\textbf{RGC \hlred{B}lock.} As shown in Fig.~\ref{rgc}, \hlred{the} RGC block contains several graph convolution\hlred{al} layers and residual skip-connections, which is inspired from \cite{wu2019point}. It takes \hlred{the} feature tensor $F_{in}$ as input\hlred{ }and outputs $F_{out}$ of the same shape $n\times c$ as $F_{in}$. 

The graph convolution in the RGC block is defined on a graph $\mathrm{G}=(V, \varepsilon)$, where \sq{$V$} denotes the node set and $\varepsilon$ denotes \hlred{the} corresponding adjacency matrix. The graph convolution is formulated as \hlred{follows}:

\begin{equation}
\begin{aligned}
f_{out}^{p}&={\omega}_0 \ast  f_{in}^{p}+{\omega}_1 \ast \Sigma_{q \in N(p)} f_{in}^{q}, \forall p \in V\\
\end{aligned}
\end{equation}
where $f_{in}^{p}$ denotes the input feature of vertex $p$, \hlred{and} $f_{out}^{p}$ represents \hlred{the} output feature of vertex $p$ after graph convolution,\hlred{ }where ${\omega}$ is the learn-able parameters and $\ast$ denotes the point-wise convolution\hlred{al} operation.

The core idea of the RGC block is to separately operate \hlred{the} convolution on the central-point feature and\hlred{ }neighbor feature, as \hlred{illustrated} in Fig. \ref{metargc}. For the neighbor features, they are grouped with the $k$ nearest neighbors of the input point cloud $x$ and then go through \hlred{a} $1\times 1$ graph convolution. \hlred{The central-point features} are convolved separately from \hlred{those of} the \hlred{neighbors }and then \hlred{are} concatenated with \hlred{the} neighbors' features. Moreover, residual skip-connections are introduced \hlred{to address} the vanishing gradient and slow convergence problems. In our implementation, we set $k=8, c=128$, and a total of 22 RGC \hlred{b}locks are used. Among them, the second one is a special meta-RGC block, which \hlred{is} described in detail next.


\begin{figure*}
\begin{center}
\includegraphics[width=0.7\textwidth]{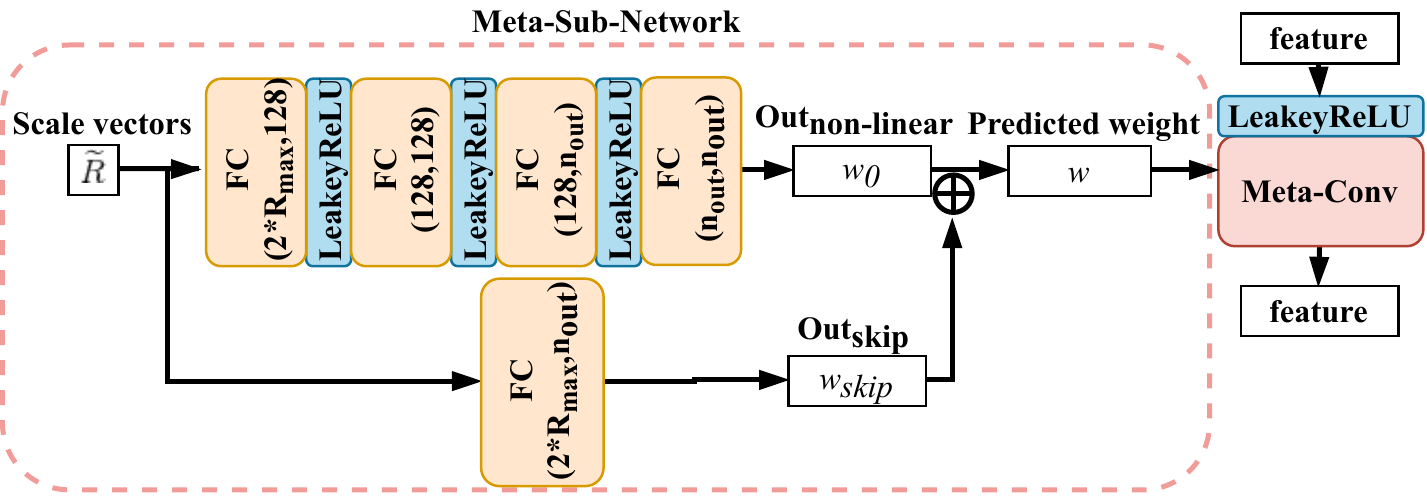}
\end{center}
   \caption{Structure of \hlred{the} \hlred{meta-subnetwork}. \hlred{The meta-subnetwork} inside the pink box predicts weights for convolution\hlred{al} layers in the \hlred{meta-RGC block}.}
\label{metaconv}

\end{figure*}

\vspace{1em}
\noindent\textbf{Meta-RGC \hlred{B}lock and \hlred{Meta-subnetwork}.} To solve the arbitrary-scale point cloud upsampling problem with a single model, we propose a meta-RGC block, which is the core part of\hlred{ }Meta-PU. The meta-RGC block is similar to a normal RGC block, but the graph convolution\hlred{al} weights are dynamically predicted, depending on the given scale factor $R$. Instead of feeding $R$ directly into the meta-RGC block, we create a scale vector $\widetilde R=\{max(0,R-i),R\}_{i=1... \lceil R \rceil)}$ as the input and fill the rest with $\{-1,-1\}$ to achieve the size $2*R_{max}$. The philosophy behind this design is indeed inspired by \hlred{m}eta-SR \cite{hu2019meta}. More specifically, \hlred{because} each input point is essentially transformed into a group of output points, $\{max(0,R-i),R\}$ can serve as the location identifier to guide the point processing network to differentiate the new $i-$th point from other points generated by the same input seed point. 

The meta convolution is formulated as \hlred{follows}:

\begin{equation}
\begin{aligned}
f_{out}^{p}&={\varphi}\left(\widetilde R ; {\theta}_0\right) \ast  f_{in}^{p}+{\varphi}\left(\widetilde R ; {\theta}_1\right) \ast \Sigma_{q \in N(p)} f_{in}^{q}, \forall p \in V\\
\end{aligned}
\end{equation}
where the convolution weights are predicted by the \hlred{meta-subnetwork} ${\varphi}(.)$ taking the scale vector $\widetilde R$ as input. Please note we have two branches of meta\hlred{-}convolution\hlred{,} as shown in Fig. \ref{metargc}. One branch is for the feature of \hlred{the} center point $p$, and \hlred{the other} is for the feature of the neighbors defined by the adjacency matrix $\varepsilon$. Since there is no pre-defined adjacency matrix $\varepsilon$ for point clouds, we define it as $N(p)$, the $k$ nearest neighbors of $p$. The convolution weights of these two branches are generated by two \hlred{meta-subnetworks} with the parameters ${\theta}_i$ respectively. 

Each \hlred{meta-subnetwork} for meta-convolution \hlred{comprises} five fully-connected (FC) layers~\cite{learn2learn2019} and several activation layers as shown in Fig.\ref{metaconv}. In the forward pass, the first FC layer takes the scale vector created from $R$ as input, and obtains a vector of $c_{hidden}$ entries. After \hlred{the} activation function, the second FC layer \hlred{produces output of} the same size as the input. Following \hlred{the} activation function, the input of the third FC layer is the $c_{hidden}$-entry encoding, and it\hlred{s} output\hlred{ has} length $c_{in} \times c_{out} \times l \times l$. Next, the fourth FC layer outputs a vector $w_{0}$ with the same shape as its input. \hlred{Unlike} the previous four concatenated layers, the last FC layer serves as \hlred{a} skip-connection that obtains output $w_{skip}$ of shape $c_{in} \times c_{out} \times l \times l$ directly from $2*R_{max}$. The two outputs $w_{0}, w_{skip}$ are added and then reshaped to $(c_{in}, c_{out}, l, l)$ as the weight matrix $w$ for meta-convolution. We set  $c_{out}=c_{in}=128$ and $c_{hidden}=128$. The $l$ represents the kernel size of \hlred{the} convolution, which is set to $1$ in our implementation. In the backward pass, instead of directly updating the weight matrix of \hlred{the} convolution, we\hlred{ }calculate the gradients of \hlred{the meta-subnetwork}, \hlred{with respect to} the weights of \hlred{the} FC layers. The gradient of \hlred{the meta-subnetwork} can be naturally calculated by the Chain Rule\hlred{ to} be trained end-to-end.

The \hlred{meta-RGC block} with dynamic weights predicted by the \hlred{meta-subnetwork} is necessary for the arbitrary-scale upsampling task because the upsampled point cloud $iR$-th to $(i+1)R$-th points are generated directly based on the features of \hlred{the} $i$-th input point and its nearest neighbors extracted via \hlred{RGC }blocks. \hlred{The point locations in the output of different scale factors have to be different, to ensure that the uniformity of the upsampled points can cover the underlying surface.} \hlred{Therefore,} the embedding features \hlred{must} be adaptively adjusted according to the scale factors. Therefore adaptive adjustment of the embedding features according to the scale factor is necessary. This \hlred{adjustment} is much better than mere upsampling to $R_{max}$ times and then \hlred{performing} the downsample. \hlred{The experiments} in Section \ref{exp2} are designed to prove this.
    
\vspace{1em}
\noindent\textbf{\hlred{The u}npooling block} takes point cloud $X$ and\hlred{ }corresponding features $F_{in}$ as input\hlred{.} It is a\hlred{n} RGC-based structure, while the output channels of \hlred{the} convolution\hlred{al} layers are set to $R_{max}\times 3$. Specifically, for feature $F_{in}$ of shape $n\times c$, it is transformed to a tensor of size $n\times (R_{max}\times 3)$, subsequently reshaped to $n\times R_{max}\times 3$, \hlred{de}noted as $T_{out}$. As a residual block, similar to the residual connection of \hlred{the} input and output features in \hlred{the} RGC block, we introduce a skip connection between points. Thus\hlred{,} the tensor $T_{out}$ is then point-wisely added to $X$ to produce the output $Y'_{max}$ of shape $n\times R_{max}\times 3$. Note that the ``add'' operation\hlred{ }naturally expand\hlred{s} $x$ to $R_{max}$ copies in a broadcast manner.
    
\vspace{1em}
\noindent\textbf{\hlred{The farthest sampling block}} \hlred{performs} a farthest sampling strategy to retain $Y'$ with $n\times R$ points from $Y'_{max}$ with $\left \lfloor R_{max}\times n \right \rfloor$. The advantages\hlred{ }are two-fold. First, the farthest sampling can sample an arbitrary number of points from the input point set, which helps obtain the required number of points as output. Second, since the farthest sampling iteratively constructs a point set with \hlred{the} farthest point-wise distance according to the \sq{Euclidean distance} from a global perspective, this step further enhances the uniformity of the point set distribution.

\subsection{Loss Function}\label{324}

For end-to-end training of\hlred{ }Meta-PU, we adopt a compound loss with both reconstruction terms $\mathcal{L}_{\mathrm{rec}}$ and uniform terms $\mathcal{L}_{\mathrm{uni}}, \mathcal{L}_{\mathrm{rep}}$:

\begin{equation}
\mathcal{L}=\lambda_{\mathrm{rec}} \mathcal{L}_{\mathrm{rec}}+\lambda_{\mathrm{uni}} \mathcal{L}_{\mathrm{uni}}+\lambda_{\mathrm{rep}} \mathcal{L}_{\mathrm{rep}}
\end{equation}
The latter two terms aim at encouraging the uniformity of \hlred{the} generated point cloud and improving \hlred{the} visual quality.

\vspace{1em}
\noindent\textbf{Repulsion loss} \cite{yu2018pu} $\mathcal{L}_{\mathrm{rep}}$ is represented as \hlred{follows}:
    
\begin{equation}
\mathcal{L}_{rep}=\sum_{i=0}^{N} \sum_{i^{\prime} \in K(i)} \eta\left(\left\|p_{i^{\prime}}-p_{i}\right\|\right) w\left(\left\|p_{i^{\prime}}-p_{i}\right\|\right)
\end{equation}
where $N$ is the number of output points, $K(i)$ is the index set of the k-nearest neighbors of point $p_i$ in output point cloud $Y'$, $\eta(r)=-r$ is the repulsion term, and $w(r)=e^{-r^{2} / h^{2}}$.

\vspace{1em}
\noindent\textbf{Uniform loss}\hlred{ The term} \cite{Li_2019_ICCV} $\mathcal{L}_{\mathrm{uni}}$ \hlred{comprises} two parts: $U_{\text {imbalance }}$ accounting for global uniformity\hlred{,} and $U_{\text {clutter }}$ accounting for the local uniformity.

\begin{equation}
\mathcal{L}_{\mathrm{uni}}=\sum_{j=1}^{M} U_{\mathrm{imbalance}}\left(S_{j}\right) \cdot U_{\mathrm{clutter}}\left(S_{j}\right)
\end{equation}
where $S_{j}, j=1..M$ refers to the ball queried point subsets with radius $r_d$ and centered at $M$ seed points farthest sampled from $Y'$. 

$$
U_{\text {imbalance }}\left(S_{j}\right)=\frac{\left(\left|S_{j}\right|-\hat{n}\right)^{2}}{\hat{n}}
$$
where $\hat{n}=\hat{N}\times r^2_d$, referring to the expected number of points in $S_{j}$. \sq{Note that, the imbalance term is not differentiable, which acts as a weight for the following clutter term.}
$$
U_{\text {clutter }}\left(S_{j}\right)=\sum_{k=1}^{\left|S_{j}\right|} \frac{\left(d_{j, k}-\hat{d}\right)^{2}}{\hat{d}}
$$
where $d_{j, k}$ is \hlred{the} point-to-neighbor distance of the $k$-th point in $S_{j}$, while $\hat{d}=\sqrt{\frac{2 \pi r_{d}^{2}}{\left|S_{j}\right| \sqrt{3}}}$ denotes the expected distance.

\vspace{1em}
\noindent\textbf{Un-biased Sinkhorn divergences} \cite{feydy2019interpolating} \hlred{are} proposed by us as \hlred{the} reconstruction loss, to encourage the distribution of generated points to lie on the underlying mesh surface. It is the interpolation between \hlred{the} Wasserstein distance\hlred{ }and kernel distance. The Sinkhorn divergences between output $Y'$ and the groundtruth $Y$  can be formulated as \hlred{follows}:

\begin{equation}
\begin{aligned}
\mathcal{L}_{\mathrm{rec}}&=\mathrm{S}_{\varepsilon}(Y, Y')=\mathrm{OT}_{\varepsilon}(Y, Y')\\
&-\frac{1}{2} \mathrm{OT}_{\varepsilon}(Y, Y) -\frac{1}{2} \mathrm{OT}_{\varepsilon}(Y', Y')
\end{aligned}
\end{equation}
where $\varepsilon$ is the regularization parameter, and 
$$
\mathrm{OT}_{\varepsilon}(Y, Y') \stackrel{\text { def. }}{=} \min _{\pi_{1}=Y, \pi_{2}=Y'} \int_{\mathcal{X}^{2}} \mathrm{C} \mathrm{d} \pi+\varepsilon \mathrm{KL}(\pi \mid Y \otimes Y')
$$
\sq{with \hlred{a} cost function on the feature space} $\mathcal{X} \subset \mathbb{R}^{\mathrm{D}}$ of dimension $\mathrm{D}$ as follows:
\begin{equation}
C(x, y)=\frac{1}{2}\|x-y\|_{2}^{2}
\end{equation}
\sq{and where\hlred{ }optimization is performed over \hlred{the} coupling measures} $\pi \in \mathcal{M}_{1}^{+}\left(\mathcal{X}^{2}\right)$ as $\left(\pi_{1}, \pi_{2}\right)$ denotes the $\pi$'s two marginal.

\subsection{Variable-scale Training Strategy}\label{325}

In the training process of most existing single-scale point cloud upsampling methods, each model is trained with one scale factor. However, \hlred{because the} scale factor varies in our arbitrary-scale upsampling task, we need to design a variable-scale training scheme to train all factors jointly. We first sampled all factors from the range of $1.1$ to $R_{max}$ with \hlred{a} stride \hlred{of} $0.1$, and put them in\hlred{ }set $\mathcal{S}_R$. For each epoch, a scale factor, say $R$, is randomly sampled from $\mathcal{S}_R$, and this factor is shared in a batch. To avoid overfitting, we also perform a series of data augmentation: rotation, random scaling, shifting, jittering, and perturbation with low probability.

\section{Experiments}

\subsection{Datasets and Metrics}

\noindent\textbf{Dataset.} For training and testing, we utilize the same dataset adopted by  PU-Net~\cite{yu2018pu} and AR-GCN~\cite{wu2019point}. This dataset contains $60$ different models from the Visionair repository. Following the protocol in the above two works, $40$ models are used for training and the rest $20$ models are \hlred{used} for testing. 

For training, 100 patches are extracted from each model, thus we have a total of \hlred{$40,000$} patches. We uniformly sample $N$ points \hlred{using} Poisson disk sampling from each patch as \hlred{the ground truth}, and non-uniformly sample $n$ points from the \hlred{ground truth} as input, where $n = \left \lfloor n_{max}\times \frac{1}{R} \right \rfloor$, and $N = R \times \left \lfloor n_{max}\times \frac{1}{R} \right \rfloor$, corresponding to the scale factor $R$. \hlred{Moreover, we set} $n_{max}=4096$ \hlred{as} the maximum number\hlred{ }of points in our training. For testing, we use the whole model instead of the patch. The sampling process of \hlred{the} ground truth and input is similar to that in training. But constrained by the GPU memory limit, we set different numbers of input points for different scale factor\hlred{s}, i.e.  \hl{$5000$ for $R<=4$, $4000$ for $4<R<=6$, $3000$ for $6<R<=12$, and $2500$ for $R>12$.}

\begin{table*}[ht]
\centering
\footnotesize
\caption{Experiments of quantitative comparisons. Single-scale models (including AR-GCN and PU-GAN) trained with each specific scale factor (top \hlred{two} rows) vs. \hlred{the} naive approach of arbitrary-scale upsampling (\hlred{rows 3 to 5}) vs. our\hlred{ }full model\hlred{ }(last row). The NUC scores are tested with $p=0.8\%$.}
\label{table1}
\resizebox{\textwidth}{!}{%
\begin{tabular}{c|cccccc|cccccc}
\hline
\multirow{2}{*}{Methods\textbackslash{}Scales} & \multicolumn{6}{c|}{2x} & \multicolumn{6}{c}{4x}                                                                                                                                 \\ \cline{2-13} 
 & CD & EMD & F-score & NUC & mean & std & CD & EMD & F-score & NUC & mean & std                     \\ \hline
AR-GCN & - & - & - & - & - & - & 0.0086 & 0.018 & 70.09\% & 0.339 & 0.0029 & 0.0033                  \\
PU-GAN & 0.016 & 0.0090 & 32.17\% & 0.249 & 0.012 & 0.015 & 0.0097 & 0.016 & 69.75\% & 0.202 & 0.0030 & 0.0031                  \\
AR-GCN(x16) & \multirow{2}{*}{0.015} & \multirow{2}{*}{0.023} & \multirow{2}{*}{30.14\%} & \multirow{2}{*}{0.307} & \multirow{2}{*}{0.0089} & \multirow{2}{*}{0.014} & \multirow{2}{*}{0.012} & \multirow{2}{*}{0.041} & \multirow{2}{*}{45.34\%} & \multirow{2}{*}{0.256} & \multirow{2}{*}{0.0081} & \multirow{2}{*}{0.0096} \\
+random-sampling &  &  &  &  &  &  &  &  &  &  &  & \\ 
AR-GCN(x16) & \multirow{2}{*}{0.014} & \multirow{2}{*}{0.012} & \multirow{2}{*}{33.52\%} & \multirow{2}{*}{0.227} & \multirow{2}{*}{0.0088} & \multirow{2}{*}{0.011} & \multirow{2}{*}{0.011} & \multirow{2}{*}{0.018} & \multirow{2}{*}{52.67\%} & \multirow{2}{*}{0.318} & \multirow{2}{*}{0.0072} & \multirow{2}{*}{0.0092} \\
+farthest-sampling &  &  &  &  &  &  &  &  &  &  &  & \\ 
AR-GCN(x16) & \multirow{2}{*}{0.015} & \multirow{2}{*}{0.013} & \multirow{2}{*}{36.98\%} & \multirow{2}{*}{0.273} & \multirow{2}{*}{0.0067} & \multirow{2}{*}{0.0082} & \multirow{2}{*}{0.013} & \multirow{2}{*}{0.013} & \multirow{2}{*}{54.05\%} & \multirow{2}{*}{0.288} & \multirow{2}{*}{0.0066} & \multirow{2}{*}{0.0080} \\
+disk-sampling &  &  &  &  &  &  &  &  &  &  &  & \\ \hline
ours & \textbf{0.010} & \textbf{0.0049} & \textbf{53.20\%} & \textbf{0.127} & \textbf{0.0023} & \textbf{0.0029} & \textbf{0.0080} & \textbf{ 0.0078 } & \textbf{ 74.05\%} & \textbf{0.192 } & \textbf{0.0022 } & \textbf{ 0.0027 }                 \\ \hline
\end{tabular}
}
\bigskip\\
\resizebox{\textwidth}{!}{%
\begin{tabular}{c|cccccc|cccccc|cccccc}
\hline
\multirow{2}{*}{Methods\textbackslash{}Scales} & \multicolumn{6}{c|}{6x} & \multicolumn{6}{c|}{9x} & \multicolumn{6}{c}{\hl{16x}}                  \\ \cline{2-19} 
 & CD & EMD & F-score & NUC & mean & std & CD & EMD & F-score & NUC & mean & std & CD & EMD & F-score & NUC & mean & std                    \\ \hline
AR-GCN & - & - & - & - & - & - & \textbf{ 0.0081} & 0.022 & \textbf{74.63\%  } & 0.344 & 0.0034 & 0.0044 & 0.0085 &	0.023 &	75.33 &	0.522 &	0.0029 &	0.0037
                 \\
PU-GAN & 0.012 & \textbf{0.013 } & 58.56\% & 0.287 & 0.011 & 0.018 & 0.0091 & \textbf{0.0085} & 70.61\% & \textbf{0.212 } & 0.0047 & 0.0057 & 0.0092 &	0.022 &	70.79 &		0.431 & 0.0042	& 0.0041
                \\
AR-GCN(x16) & \multirow{2}{*}{0.013} & \multirow{2}{*}{0.025} & \multirow{2}{*}{46.11\%} & \multirow{2}{*}{0.309} & \multirow{2}{*}{0.0077} & \multirow{2}{*}{0.0099} & \multirow{2}{*}{0.011} & \multirow{2}{*}{0.039} & \multirow{2}{*}{48.80\%} & \multirow{2}{*}{0.417} & \multirow{2}{*}{0.0080} & \multirow{2}{*}{0.0086}  & - & - & - & - & - & - \\
+random-sampling &  &  &  &  &  &  &  &  &  &  &  & \\ 
AR-GCN(x16) & \multirow{2}{*}{0.011} & \multirow{2}{*}{0.078} & \multirow{2}{*}{53.97\%} & \multirow{2}{*}{2.808} & \multirow{2}{*}{0.0075} & \multirow{2}{*}{0.0085} & \multirow{2}{*}{0.011} & \multirow{2}{*}{0.015} & \multirow{2}{*}{53.89\%} & \multirow{2}{*}{5.522} & \multirow{2}{*}{0.0081} & \multirow{2}{*}{0.0081}  & - & - & - & - & - & - \\
+farthest-sampling &  &  &  &  &  &  &  &  &  &  &  & \\ 
AR-GCN(x16) & \multirow{2}{*}{0.012} & \multirow{2}{*}{0.014} & \multirow{2}{*}{59.41\%} & \multirow{2}{*}{0.293} & \multirow{2}{*}{0.0065} & \multirow{2}{*}{0.0079} & \multirow{2}{*}{0.011} & \multirow{2}{*}{0.014} & \multirow{2}{*}{62.70\%} & \multirow{2}{*}{0.298} & \multirow{2}{*}{0.0065} & \multirow{2}{*}{0.0078}   & - & - & - & - & - & - \\
+disk-sampling &  &  &  &  &  &  &  &  &  &  &  & \\ \hline
ours & \textbf{ 0.0083 } & 0.014 & \textbf{74.98\% } & \textbf{ 0.267 } & \textbf{0.0025} & \textbf{0.0030} & 0.0083 & 0.016 & 73.74\% & 0.274 & \textbf{0.0030  } & \textbf{0.0034  }  & 0.0082 &	0.023 &	75.62 &		0.428 &	0.0029	& 0.0035
             \\ \hline
\end{tabular}
}
\end{table*}

\begin{table}[ht]
\begin{center}
\setlength{\tabcolsep}{1.0mm}{
\caption{Quantitative comparisons with \hlred{the} EAR. The NUC scores are tested with $p=0.8\%$.}
\label{table: EAR}
\resizebox{0.45\textwidth}{!}{%
\begin{tabular}{ccccccc}
\toprule
            Methods & CD$\downarrow$ & EMD$\downarrow$ & F-score$\uparrow$ & NUC$\downarrow$ & mean$\downarrow$ & std$\downarrow$ \\ \hline
EAR (2x) & 0.0113 & 0.0214 & 48.07\% & 0.747 & 0.0048 & 0.0113          \\
Ours (2x) & \textbf{0.010} & \textbf{0.0049} & \textbf{53.20\%} & \textbf{0.127} & \textbf{0.0023} & \textbf{0.0029}          \\
EAR (4x) & 0.0112 & 0.0176 & 51.26\% & 0.478 & 0.0074 & 0.0137          \\
Ours (4x) & \textbf{0.0080} & \textbf{0.0078} & \textbf{74.05\%} & \textbf{0.192} & \textbf{0.0022} & \textbf{0.0027}          \\
EAR (6x) & 0.0120 & 0.0184 & 52.26\% & 0.421 & 0.0085 & 0.0145          \\
Ours (6x) & \textbf{0.0083} & \textbf{0.014} & \textbf{74.98\%} & \textbf{0.267} & \textbf{0.0025} & \textbf{0.0030}          \\
EAR (9x) & 0.0119 & 0.0174 & 52.93\% & 0.442 & 0.0089 & 0.0140          \\
Ours (9x) & \textbf{0.0083} & \textbf{0.016} & \textbf{73.74\%} & \textbf{0.274} & \textbf{0.0030} & \textbf{0.0034}          \\ \bottomrule
\end{tabular}
}
}
\end{center}
\end{table}

\vspace{1em}
\noindent\textbf{Metrics.} For \hlred{a} fair comparison, we employ several different popular metrics: \textbf{\textit{Chamfer Distance (CD)}} and \textbf{\textit{Earth Mover Distance (EMD)}} defined on the \hlred{E}uclidean distance are to measure the difference between predicted points $Y'$ and ground-truth point cloud $Y$. \hlred{The }CD sums the square of the distance between each point and the nearest point in the other point set, then calculates the average for each point set. \hlred{The }EMD measures the minimum cost of turning one of the point sets into the other. For these two metrics, the lower, the better. We also report \hlred{the} \textbf{\textit{F-score}} between $Y'$ and $Y$ that defines the point cloud super-resolution as a classification problem as \cite{wu2019point}. For this metric, larger is better. We employ \hlred{the} \textbf{\textit{normalized uniformity coefficient (NUC)}} ~\cite{yu2018pu} to evaluate the uniformity of $Y'$ by directly comparing the output point cloud $Y'$ with corresponding ground\hlred{-}truth meshes, and \textbf{\textit{deviation mean and std}} to measure the difference between the output point cloud and\hlred{ }\hlred{ground-truth} mesh. For these two metrics, smaller is better.


%

\subsection{Implementation Details}
We train the network for 60 epoch\hlred{s} with a batch size of 18. Adam is adopted as the loss optimizer. The learning rate is initially set to 0.001 for \hlred{FC} layers and 0.0001 for convolutions and other parameters, which is decayed with a cosine annealing scheduler to $1e-5$. Parameters $\lambda_{\mathrm{rec}}$, $\lambda_{\mathrm{uni}}$ and $\lambda_{\mathrm{rep}}$ for the joint loss function are set to $1$, $0.001$ and $0.005$ respectively. Generally, the training takes less than \hlred{seven hours} on two Titan-XP GPUs.
 Theoretically,\hlred{ }MetaPU supports any large scale, but we set the max\hlred{imum} scale to 16 due to \hlred{the} limitations of \hlred{the} computing resources and practical needs.


\begin{table*}[ht]
\caption{\hl{Quantitative comparisons with MPU. Our method obtains superior results under most metrics.}}
\centering
\begin{tabular}{@{}cccccccccccc@{}}
\toprule
\multirow{2}{*}{Method} & \multirow{2}{*}{CD} & \multirow{2}{*}{EMD} & \multirow{2}{*}{F-score} & \multicolumn{5}{c}{NUC with different p} & \multicolumn{2}{c}{Deviation(1e-2)} & \multirow{2}{*}{Time} \\ \cmidrule(r){5-11} 
 &  &  &  & 0.2\% & 0.4\% & 0.6\% & 0.8\% & 1.0\% & mean & std  &             \\ \cmidrule(r){1-12}
MPU(2x) & \textbf{0.0097} &	0.0132 &	52.65\% &	0.317 &	0.271 &	0.252 &	0.241 &	0.234 &	\textbf{0.21} &	0.30 &	5.16
         \\
ours(2x) & 0.010 & \textbf{0.0049} & \textbf{53.20\%} & \textbf{0.183} & \textbf{0.147} & \textbf{0.134} & \textbf{0.127} & \textbf{0.123} & 0.23 & \textbf{0.29}  & \textbf{0.90}       \\
MPU(4x) & 0.0086 &	0.012 &	73.16\% &	0.321 &	0.282 &	0.265 &	0.256 &	0.249 &	\textbf{0.22} &	0.28 &	36.28
         \\
ours(4x) & \textbf{0.0080} & \textbf{0.0078} & \textbf{74.05\%} & \textbf{0.245} & \textbf{0.213} & \textbf{0.200} & \textbf{0.192} & \textbf{0.187} & \textbf{0.22} & \textbf{0.27}  & \textbf{0.91}       \\
MPU(16x) & \textbf{0.0078} &	\textbf{0.023} &	\textbf{76.35\%} &	\textbf{0.425} &	\textbf{0.378} &	\textbf{0.357} &	\textbf{0.346} &	\textbf{0.337} &	\textbf{0.28} &	\textbf{0.35} &	389.40
         \\
ours(16x) & 0.0082 & \textbf{0.023} & 75.62\% & 0.555 & 0.482 & 0.448 & 0.428 & 0.414 & 0.29 & \textbf{0.35}  & \textbf{0.50}       \\
\bottomrule
\end{tabular}
\label{table: MPU}
\end{table*}

\begin{table}[ht]
\caption{Comparison of the inference time. }

\begin{center}
	\begin{tabular}{cp{2cm}<{\centering}p{2cm}<{\centering}cc}
		\toprule
		Method & AR-GCN + Disk-sampling & PU-GAN + Disk-sampling & EAR & ours    \\ \midrule
		Time(s) & 10.28 & 10.06 & 351.10 & \textbf{0.79}\\ \bottomrule
	\end{tabular}
\end{center}
\label{tab:p & t}
\end{table}

\subsection{Ablation Analysis}
\begin{figure}[ht]
\begin{center}
\includegraphics[width=0.48\textwidth]{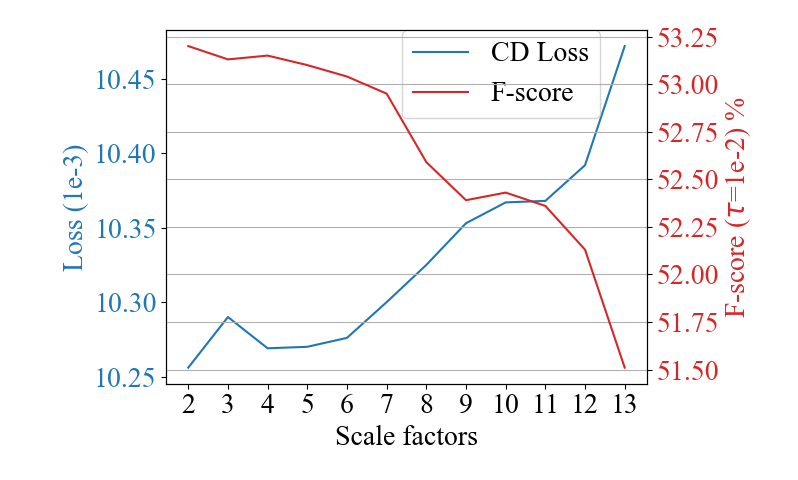}
\end{center}

   \caption{Ablation on the meta-RGC block. Meta-PU is applied to upsample the point clouds to 2x, but the weights of its meta-RGC block are generated with \hlred{a} different input scale factor $R$. $R=2$ achieves the best performance on both F-score and CD, which indicates our meta-RGC block \hlred{adapts} the convolution\hlred{al} weight appropriately to different scale factors. }
\label{meta-curve}
\end{figure}

\subsection{Comparison with Single-scale Upsampling Methods}\label{exp1}
In this experiment, we compare\hlred{ }Meta-PU with state-of-the-art single-scale upsampling methods, including PU-GAN~\cite{Li_2019_ICCV} and AR-GCN~\cite{wu2019point}, to upsample the sparse point cloud with scale factors \hl{$R\in[2,4,6,9,16]$}\hlred{.} Their models are trained with \hlred{the} author-released code, and all settings are the same as stated in their papers. Since they are single-scale upsampling methods, for each scale factor, an individual model is trained. Due to the limitation\hlred{s} of the two-stage upsampling strategy, AR-GCN can only be trained with the factors \hl{$4, 9, 16$}, \hlred{whereas} PU-GAN can be trained with all four factors. Their performance is reported in the first two rows of Table \ref{table1}. We surprisingly observe that our arbitrary-upscale model even outperforms their single-scale models with \hl{most} scale factors. Particularly, our model performs significantly better on \hlred{the} F-score, NUC, mean\hlred{,} and std metrics than other models, and is more stable on all scales. This may be because multiple joint training tasks of different scales can benefit each other, thus improving performance. In addition,\hlred{ }Meta-PU\hlred{ }needs to train \hlred{only} once for all testings, while others need to train multiple models, which is very inefficient.

\subsection{Comparison with Naive Arbitrary-scale Upsampling Approaches}\label{exp2}
    \hlred{A} naive approach to achieve arbitrary-scale upsampling is \hlred{to first} use a state-of-the-art single-scale model to upsample the cloud point to a large scale, and then downsample it to a specific smaller scale. We compare our method with this naive approach. Specifically, we choose AR-GCN~\cite{wu2019point} to upsample point clouds to 16\hlred{x} and then downsample them to 2x,4x,6x and 9x with the random sampling, disk sampling, and farthest sampling algorithms\hlred{.} \hlred{The results} are reported in 3-5th rows of Table \ref{table1}. We can see that random sampling gets the worst scores because it non-uniformly downsamples the points. In comparison, the more advanced sampling algorithms, including disk sampling and farthest sampling, perform better by considering uniformity. Our method is still superior to all of them because\hlred{ }the result of a smaller scale factor \hlred{in our method} is not simply a subset of the large-factor one. In fact, \hlred{Meta-PU} can adaptively adjust the location of the output points to fit the underlying surface better and \hlred{maintain} uniformity according to different scale factors. This will be analyzed by the ablation study of the \hlred{m}eta-RGC block in the next subsection. Moreover, compared to the strongest baseline (AR-GCN+disk-sampling), ours is 120 times faster (Table \ref{tab:p & t}), because this advanced upsampling algorithm requiring mesh reconstruction is slow. 

We also compare our method with \hlred{the} state-of-the-art optimization-based method EAR~\cite{huang2013edge}\hlred{, which} is also applicable to variable scales. The results of scale \hl{2,4,6,9,16} are provided in Table~\ref{table: EAR}. It could be found that our method yields superior results under all metrics.

\hlred{Further, we }compare \hlred{Meta-PU} with \hlred{the} state-of-the-art \hlred{multistep}\hlred{ }upsample method MPU~\cite{Yifan_2019_CVPR} that recursively upsamples a point set, which is also applicable to scales of a power of 2,\hlred{(} e.g\hlred{.}, 2,4,16\hlred{)}. The results of scale\hlred{s} 2,4,16 are provided in Table~\ref{table: MPU}. It can be found that our method obtains superior results under most metrics. In addition, we provide a comparison of inference times.  The running time of our method is much less at all scales, \hlred{demonstrating} that our method is more efficient than the recursive approach.

\subsection{Inference Time Comparison}
In Table \ref{tab:p & t}, we provide the comparison of the average inference time of all integer scales in (1,16]. Since Disk-sampling obtains the best performance as shown in Table~\ref{table1}, it is employed in other single-scale baselines for arbitrary-scale upsampling. We also compare the inference time of ours with the optimization-based method EAR. The running time of our method is much less than all the compared methods. Specifically, the speed of \hlred{a} trivial single-scale baseline is dragged down by the bottleneck of disk-sampling. \hlred{We also} calculated the inference time of our model at \hlred{the} scales \hlred{of} 2,4,6,9\hlred{,} and 16 with 2500 points. We found no difference in inference time for different scales\hlred{, demonstrating} the stability of our method in terms of inference time.

\subsection{Quantitative Metrics on Varying Scales} 
\begin{figure}[t]
\centering
\begin{subfigure}[F-score]{0.4\textwidth}
   \includegraphics[width=\textwidth]{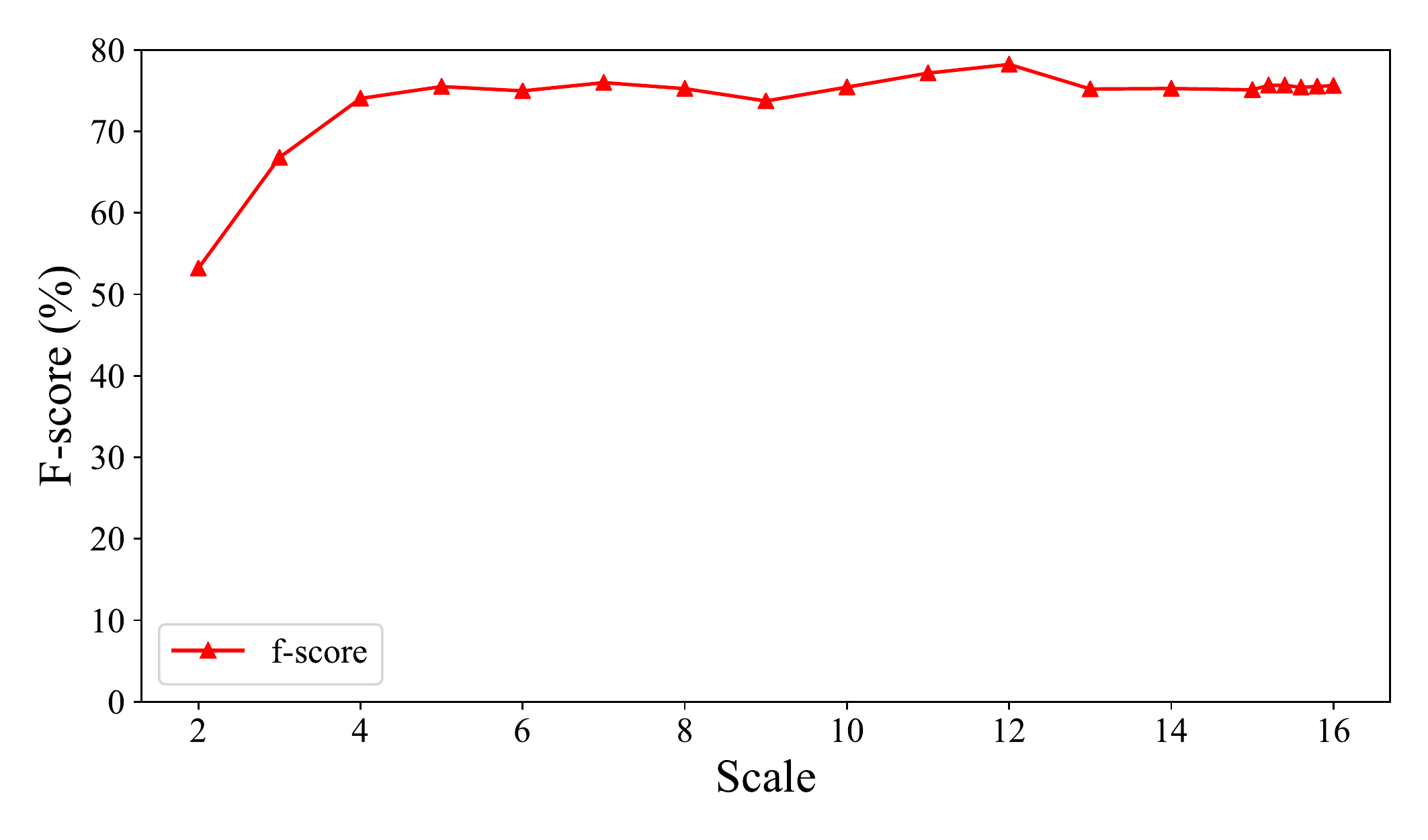}
\end{subfigure}
\begin{subfigure}[CD]{0.4\textwidth}
   \includegraphics[width=\textwidth]{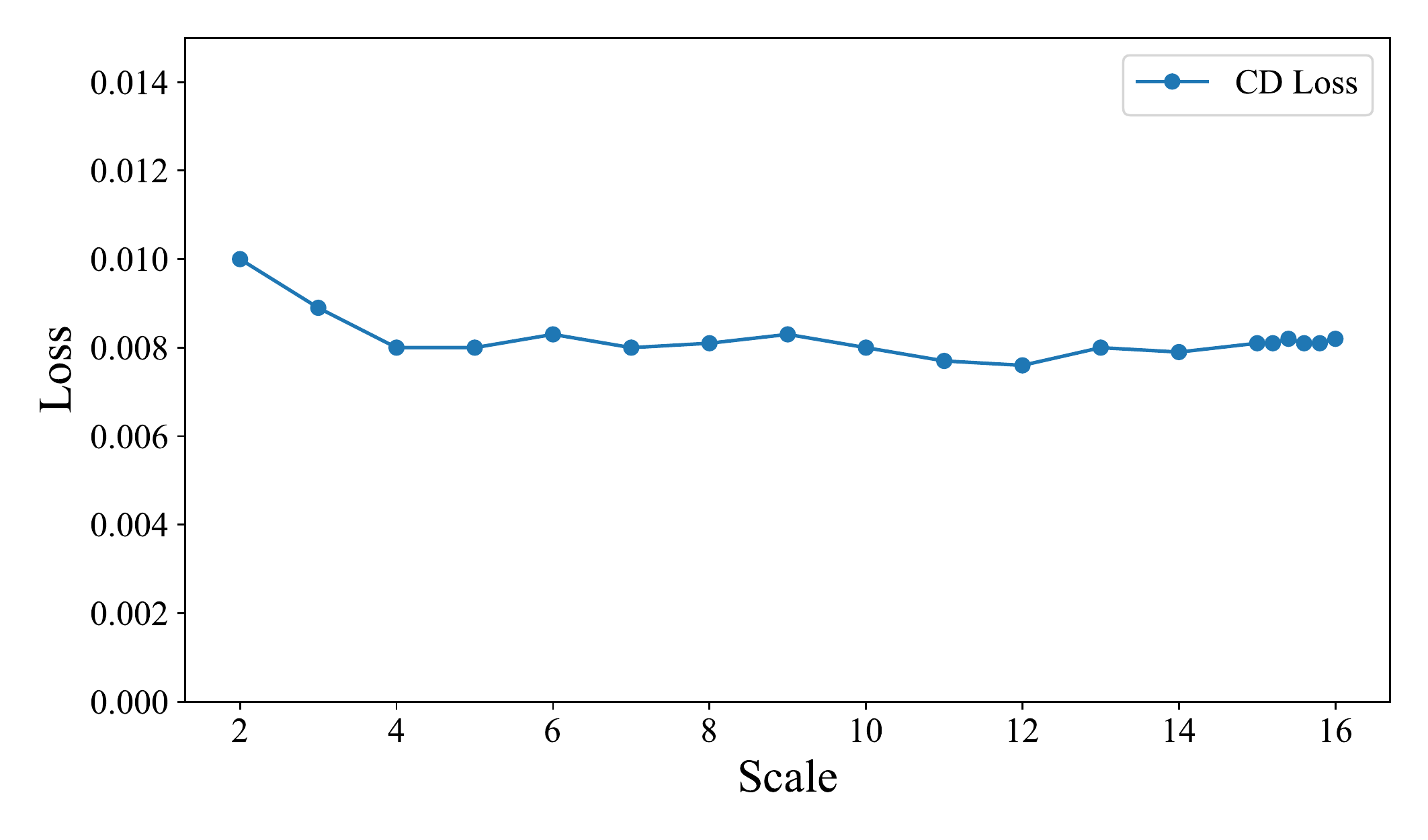}
\end{subfigure}
\caption{F-scores and CD losses of \hlred{upsampling} on different scales with Meta-PU. Our method performs stably on different scales.}
\label{curve}
\end{figure}

Fig. \ref{curve} shows the quantitative comparison results of \hlred{the} F-scores and CD loss for point cloud upsampling with different scales. It \hlred{demonstrates} that our method can support a wide range of scales, and \hlred{that} the performances on different scales are stable. We need to note that the lower F-score for small scales (e.g.,\hlred{ }2) is because one fixed distance threshold is used in calculating the F-scores, causing the F-score to be relatively low when the point cloud is too sparse.

\begin{table}[ht]
\caption{\hl{Ablation on the meta-RGC block.}}
\centering
\begin{tabular}{@{}cccc@{}}
\toprule
Methods\textbackslash{}F-score & 2x    & 4x    & 6x    \\ \midrule
Full-model                     & 53.20 & 74.05 & 74.98 \\
Replace Meta-RGC               & 52.33 & 73.08 & 74.09 \\ \bottomrule
\end{tabular}
\label{tab:repmeta}
\end{table}

\begin{figure}[ht]
\centering
\begin{subfigure}[b]{0.15\textwidth}
\textcolor{green}{\frame{\includegraphics[width=\textwidth]{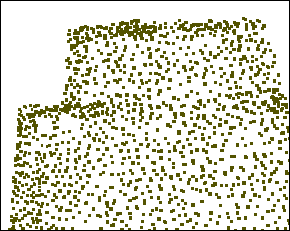}}}
\end{subfigure}
\begin{subfigure}[b]{0.15\textwidth}
\textcolor{green}{\frame{\includegraphics[width=\textwidth]{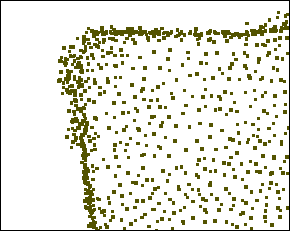}}}
\end{subfigure}
\begin{subfigure}[b]{0.15\textwidth}
\textcolor{green}{\frame{\includegraphics[width=\textwidth]{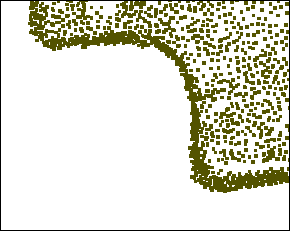}}}
\end{subfigure}
\begin{subfigure}[b]{0.15\textwidth}
\textcolor{red}{\frame{\includegraphics[width=\textwidth]{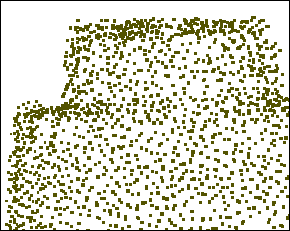}}}
\end{subfigure}
\begin{subfigure}[b]{0.15\textwidth}
\textcolor{red}{\frame{\includegraphics[width=\textwidth]{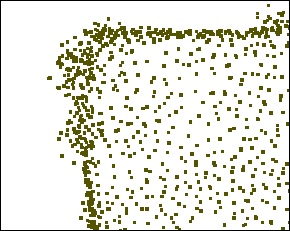}}}
\end{subfigure}
\begin{subfigure}[b]{0.15\textwidth}
\textcolor{red}{\frame{\includegraphics[width=\textwidth]{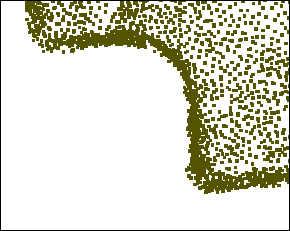}}}
\end{subfigure}
\caption{Close up of the upsampling results with and without \hlred{meta-RGC block;} {Greens}: full model, {Reds}: w/o meta-RGC block.}
\label{fig:nometa}
\end{figure}

\noindent\textbf{Importance of \hlred{the Meta-RGC Block}.} We design an experiment to evaluate the influence of the meta-RGC block quantitatively. We use\hlred{ }Meta-PU to upsample\hlred{ }point clouds to 2x, but the weights of\hlred{ the} meta-RGC block are generated with different input $R$. We measure the average F-score and CD values on the test\hlred{ing} set for different $R$, as plotted in Fig.~\ref{meta-curve}. It can be observed that the best performance \hlred{for both the F-score and CD} is achieved when the input scale factor $R$ of the meta-RGC block equals the target upsampling scale $2$. This demonstrates \hlred{the} meta-RGC block adapting the convolution weight properly to different scale factors. \hlred{Moreover}, the \hlred{m}eta-RGC block \hlred{adaptation} to various scale factors is the key to \hlred{makeing} the output points \textbf{\textit{better fit the underlying surface and keep uniform}}. To demonstrate \hlred{this}, we respectively train \hlred{the} full model and our model where \hlred{the meta-RGC block} is replaced by normal RGC-block, and show the close up of some results in Fig.\ref{fig:nometa}. It is obvious that the results of our full model are more precise, sharper and cleaner, especially around key positions\hlred{,} such as corners. Also, we \hlred{conducted an} ablation \hlred{study to replace} the \hlred{meta-RGC} with normal RGC, while keeping all the other parts fixed. The results are shown in Table~\ref{tab:repmeta}. These results demonstrate \hlred{the} meta-RGC block adapts the convolution\hlred{al} weight properly to different scale factors and improves\hlred{ }performance. Therefore, as we explained before, the performance improvements of our method mainly come from two aspects: \hlred{the} joint training of multiple scale factors with one model and the \hlred{m}eta-RGC block.



\begin{figure}[ht]
\centering
\begin{subfigure}[b]{0.2\textwidth}
\includegraphics[width=\textwidth]{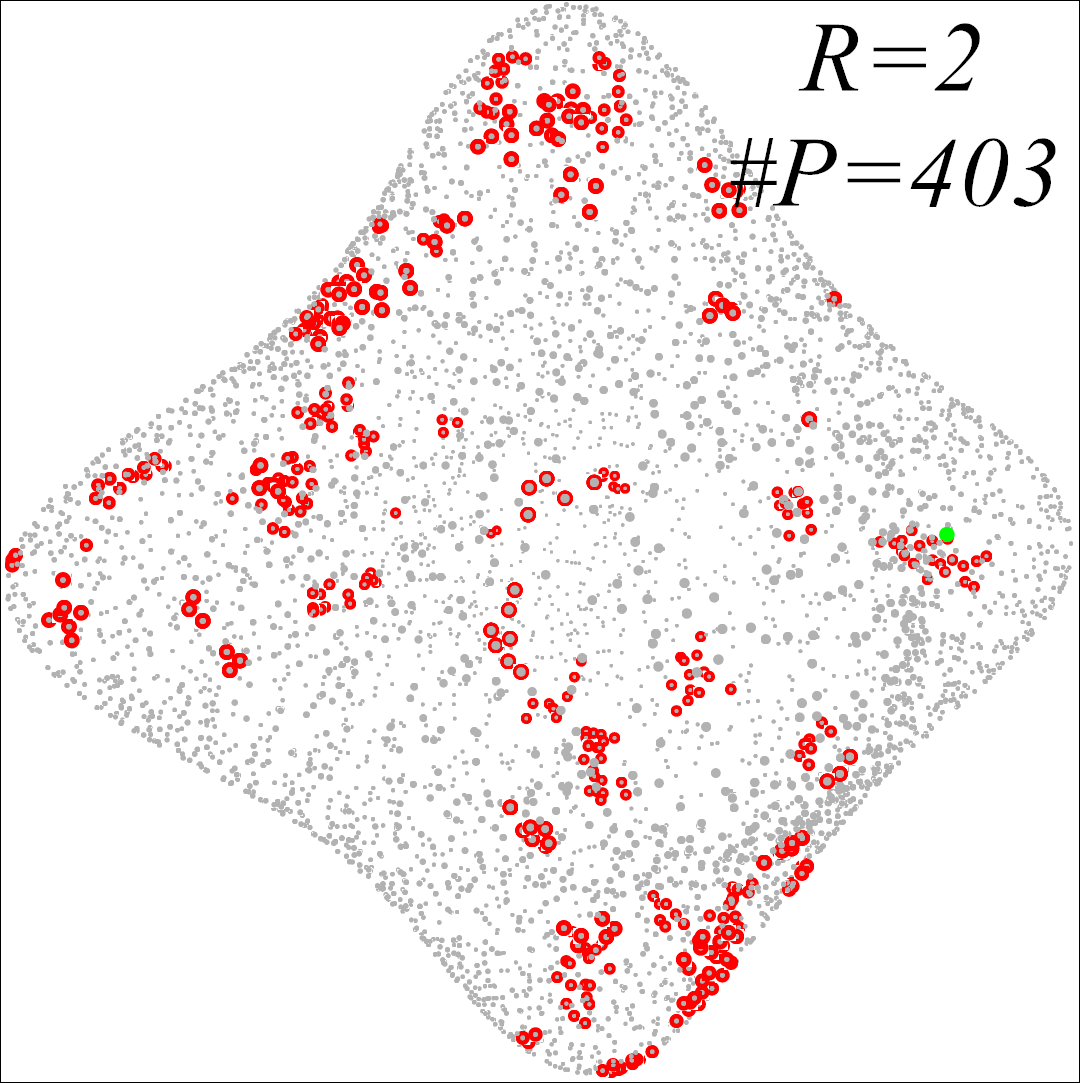}
\end{subfigure}
\begin{subfigure}[b]{0.2\textwidth}
\includegraphics[width=\textwidth]{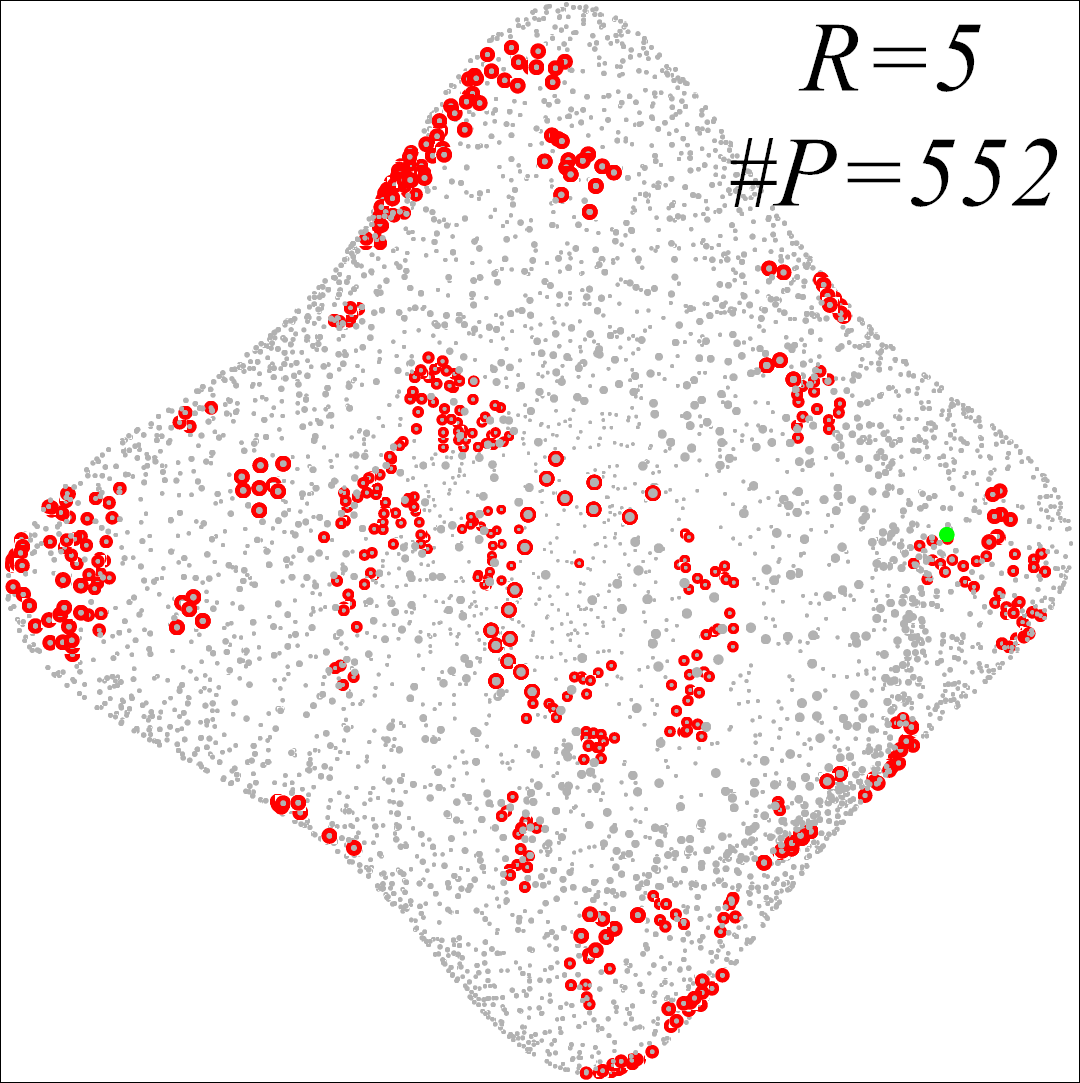}
\end{subfigure}
\begin{subfigure}[b]{0.2\textwidth}		
\includegraphics[width=\textwidth]{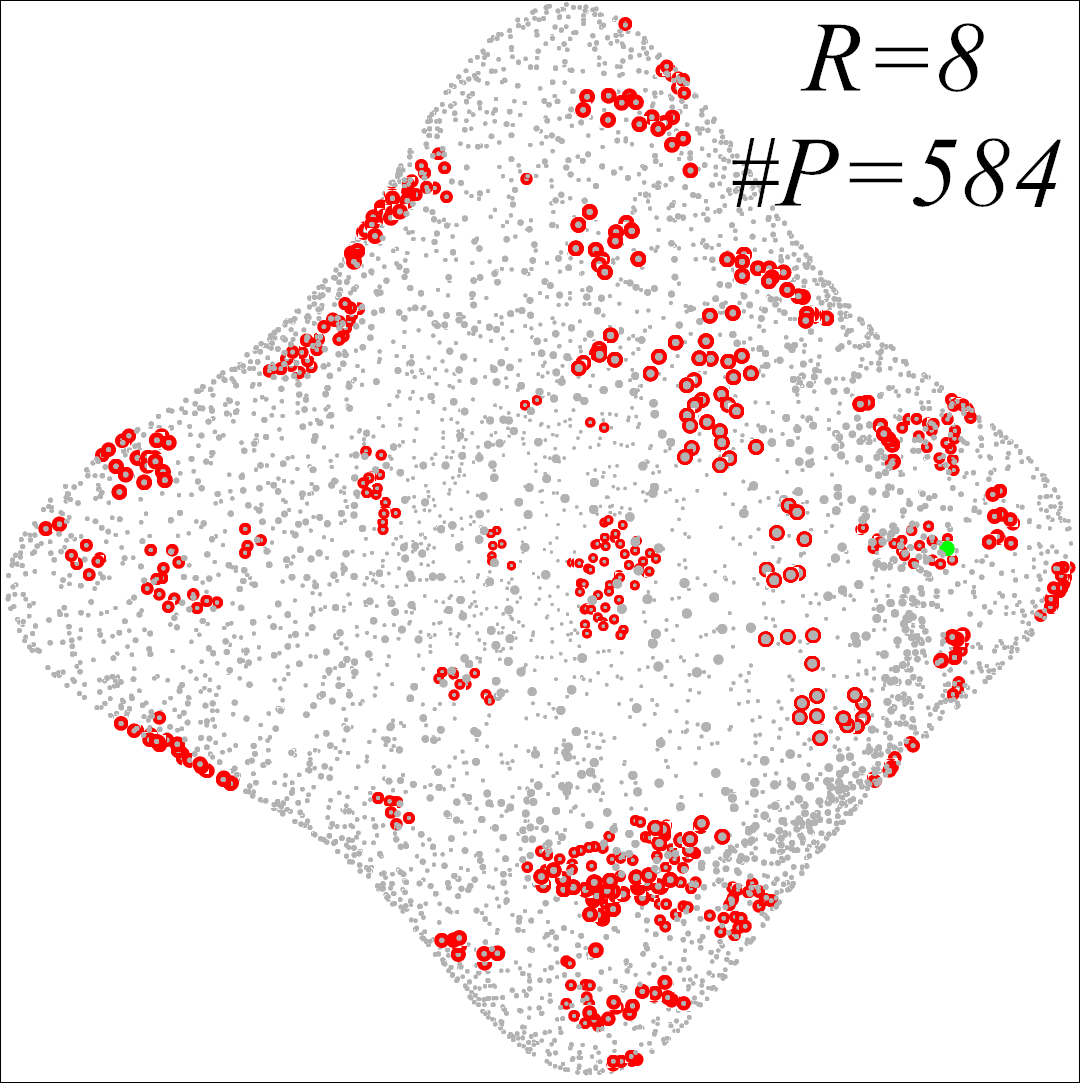}
\end{subfigure}
\begin{subfigure}[b]{0.2\textwidth}
\includegraphics[width=\textwidth]{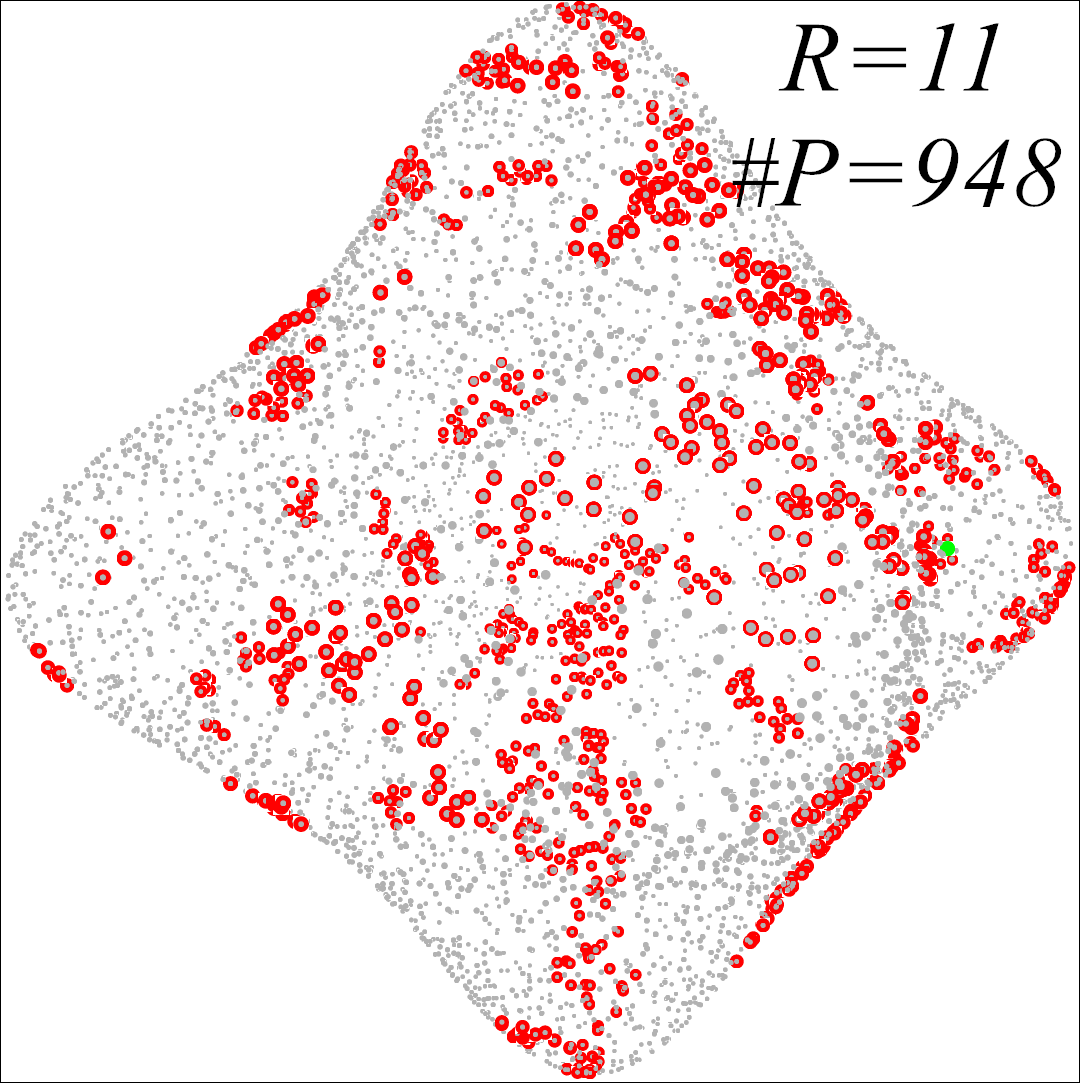}
\end{subfigure}
\caption{Visualizations of \hlred{center points} ({Green}), corresponding effective receptive fields ({Red}) in the input, and other points in the input ({Gray}). The scale factor\hlred{s are} 2,5,8,11 from left to right. $\#P$ indicates the number of input points within the receptive field.}
\label{fig:nometa2}

\end{figure}

Another important function of the \hlred{m}eta-RGC block is enabling adaptive receptive fields for different scales\hlred{, which} is useful because large-scale upsampling typically takes sparser input\hlred{ }and requires exploring long-range relationships between input points. To demonstrate this, we \hlred{analyze} the effective receptive fields. Specifically, we fix the model weights of\hlred{ }Meta-PU and remove the \hlred{f}arthest \hlred{s}ampling block to eliminate its influence. Then, we test the model with the same inputs but different scale factors. In the common input, we randomly choose a point and find its closest output point in the results of each factor as the centroids. We mask out the gradients from all points except the centroid and propagate the gradients back to the input points.
Only the input points whose gradient value\hlred{s are} larger than $1\%$ of the maximum gradient value are considered within the receptive field of the center point. As shown in Fig. \ref{fig:nometa2}, the receptive field is dynamically increased with the larger input scale factor of the \hlred{m}eta-RGC block.

\begin{table*}[ht]
\caption{Experiments on our specially designed scale tensor vs simply feed\hlred{ing} the original scale factor. All models are tested with \hlred{a} scale factor of 4. Our specially designed scale tensor performs better.}
\centering
\begin{tabular}{@{}ccccccccccc@{}}
\toprule
\multirow{2}{*}{Method} & \multirow{2}{*}{CD} & \multirow{2}{*}{EMD} & \multirow{2}{*}{F-score} & \multicolumn{5}{c}{NUC with different p} & \multicolumn{2}{c}{Deviation(1e-2)} \\ \cmidrule(r){5-11} 
 &  &  &  & 0.2\% & 0.4\% & 0.6\% & 0.8\% & 1.0\% & mean & std              \\ \cmidrule(r){1-11}
all-R & 0.0081 & \textbf{0.0078} & 73.46\% & 0.250 & 0.216 & 0.202 & 0.194 & 0.189 & 0.23 & 0.28         \\
ours & \textbf{0.0080} & \textbf{0.0078} & \textbf{74.05\%} & \textbf{0.245} & \textbf{0.213} & \textbf{0.200} & \textbf{0.192} & \textbf{0.187} & \textbf{0.22} & \textbf{0.27}         \\
\bottomrule
\end{tabular}
\label{RvsST}
\end{table*}

\vspace{1em}
\noindent\textbf{Importance of Specially Designed Scale Tensor.} In this ablation study, we \hlred{aim} to show the advantages of our designed scale tensor $\widetilde R$ with the location identifier over directly feeding $R$ into the meta-RGC block. In the first row of Table. \ref{RvsST}, we fill the scale tensor with all $R$s and train the model with the same setting. We can observe that the model with \hlred{the} specially designed scale tensor performs better\hlred{, because} the location identifier in our scale tensor contains extra information to guide the network \hlred{to} better \hlred{differentiate} a group of points, generated by the same seed point, from each other\hlred{.}

\begin{table*}[ht]
\caption{Experiments on our model with Sinkhorn loss vs. Chamfer Distance vs. GAN loss. All models are tested with \hlred{a} scale factor of 4.}
\centering
\begin{tabular}{@{}ccccccccccc@{}}
\toprule
\multirow{2}{*}{Method} & \multirow{2}{*}{CD} & \multirow{2}{*}{EMD} & \multirow{2}{*}{F-score} & \multicolumn{5}{c}{NUC with different p} & \multicolumn{2}{c}{Deviation(1e-2)} \\ \cmidrule(r){5-11} 
 &  &  &  & 0.2\% & 0.4\% & 0.6\% & 0.8\% & 1.0\% & mean & std              \\ \cmidrule(r){1-11}
ours-CD & 0.0090 & 0.0099 & 65.49\% & 0.256 & 0.230 & 0.218 & 0.211 & 0.206 & 0.47 & 0.51         \\
with GAN-loss & 0.0081	& 0.0081 &	73.69\% &	0.259 &	0.222 &	0.208 &	0.199 &	0.192 &	0.22 &	0.28         \\
ours & \textbf{0.0080} & \textbf{0.0078} & \textbf{74.05\%} & \textbf{0.245} & \textbf{0.213} & \textbf{0.200} & \textbf{0.192} & \textbf{0.187} & \textbf{0.22} & \textbf{0.27}         \\
\bottomrule
\end{tabular}
\label{CDvsWD}
\end{table*}

\vspace{1em}
\noindent\textbf{Ablation on Loss Terms.}
To validate our choice on the loss terms, we design two ablation experiments. In the first experiment, we replace \hlred{the} Sinkhorn loss with \hlred{the} CD distance in our method as \cite{wu2019point} and report the performance (ours-CD) in Table \ref{CDvsWD}. It clearly \hlred{demonstrates} the superiority of the Sinkhorn divergence reconstruction loss over \hlred{CD}. For example, the F-score using \hlred{the Sinkhorn} loss is about 7\% better than that using \hlred{the} CD loss. Rather than just measuring the distance between every nearest point between two point sets in \hlred{the} CD loss, \hlred{the} Sinkhorn loss considers the joint probability between two point sets and encourages the distribution of \hlred{the} generated points to lie on the \hlred{ground-truth} mesh surface. In the second experiment, we add \hlred{the} GAN-loss to our full-model (with GAN-loss). The scores on 4x are reported in Table \ref{CDvsWD}. We found the GAN loss does not further improve the performance of our model. Thus\hlred{,} we do not include it in our implementation.

\begin{table*}[ht]
\caption{Experiments on training\hlred{ }Meta-PU with different upscale ranges $R_{max}$ and testing with scale factor $4$. Training with a wider range brings performance improvements.}
\label{9vs16}
\centering
\begin{tabular}{@{}ccccccccccc@{}}
\toprule
\multirow{2}{*}{Method} & \multirow{2}{*}{CD} & \multirow{2}{*}{EMD} & \multirow{2}{*}{F-score} & \multicolumn{5}{c}{NUC with different p} & \multicolumn{2}{c}{Deviation(1e-2)} \\ \cmidrule(r){5-11} 
 &  &  &  & 0.2\% & 0.4\% & 0.6\% & 0.8\% & 1.0\% & mean & std \\ \cmidrule(r){1-11}
ours(max5) & \textbf{0.0079} & 0.015 & 72.85\% & 0.386 & 0.341 & 0.324 & 0.314 & 0.305 & \textbf{0.20} & \textbf{0.22}           \\
ours(max9) & 0.0081 & 0.011 & 73.39\% & 0.248 & 0.214 & 0.202 & 0.193 & 0.189 & 0.22 & 0.27         \\
ours(max16) & 0.0080 & \textbf{0.0078} & \textbf{74.05\%} & \textbf{0.245} & \textbf{0.213} & \textbf{0.200} & \textbf{0.192} & \textbf{0.187} & 0.22 & 0.27         \\
\bottomrule
\end{tabular}

\end{table*}

\vspace{1em}
\noindent\textbf{Ablation on Different Scale Range\hlred{s.}}
To test the influence of \hlred{the} scale factor ranges in our method, we design an experiment comparing our model trained with different $R_{max}$ but all tested with $R=4$. The result is reported in Table \ref{9vs16}, in which we compare the results of \hlred{the} models trained with $R_{max}=5,9,16$. Generally, ours(max16) trained with the widest range performs the best, while ours(max9) are better than ours(max5). 
From this, we can conclude that upsampling tasks with scale factors have some shared properties that allow them to benefit from each other during joint training, further allowing the models to learn some common knowledge about upsampling as well.


\begin{table}[!ht]
\begin{center}
\setlength{\tabcolsep}{1.0mm}{
\caption{Quantitative results on SHREC15 with scale=4. The NUC scores are tested with $p=0.8\%$.}
\label{table: SHREC}
\resizebox{0.45\textwidth}{!}{%
\begin{tabular}{ccccccc}
\toprule
            Methods & CD$\downarrow$ & EMD$\downarrow$ & F-score$\uparrow$ & NUC$\downarrow$ & mean$\downarrow$ & std$\downarrow$ \\ \midrule
AR-GCN & 0.0048 & 0.0127 & 94.04\% & 0.364 & 0.0018 & 0.0022          \\
Ours & \textbf{0.0045} & \textbf{0.0071} & \textbf{94.96\%} & \textbf{0.197} & \textbf{0.0015} & \textbf{0.0016}          \\ \bottomrule
\end{tabular}
}
}
\end{center}
\end{table}

\subsection{Applications}
In this section, we \hlred{reveal} the advantages of our method in different practical applications, including mesh reconstruction, point cloud classification, upsampling on real data, and upsampling with arbitrary input sizes or scales. 



\begin{figure*}[ht]
\centering
\begin{subfigure}[b]{0.18\textwidth}
\includegraphics[width=\textwidth]{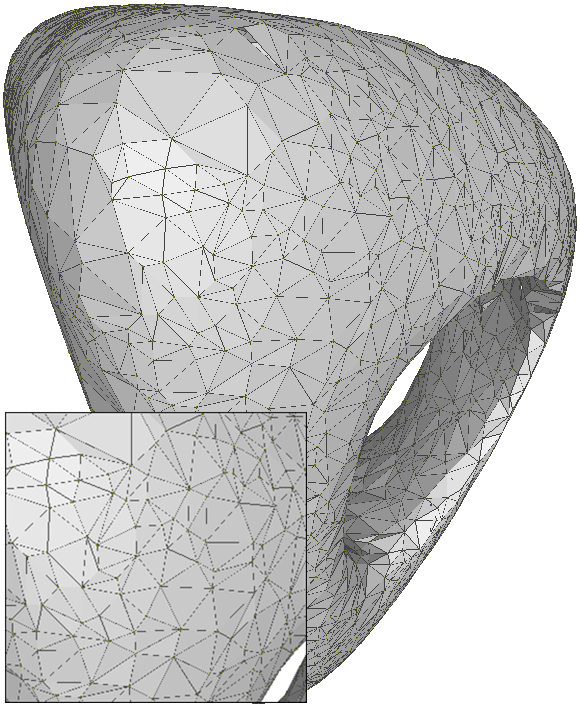}
\end{subfigure}
\begin{subfigure}[b]{0.18\textwidth}
\includegraphics[width=\textwidth]{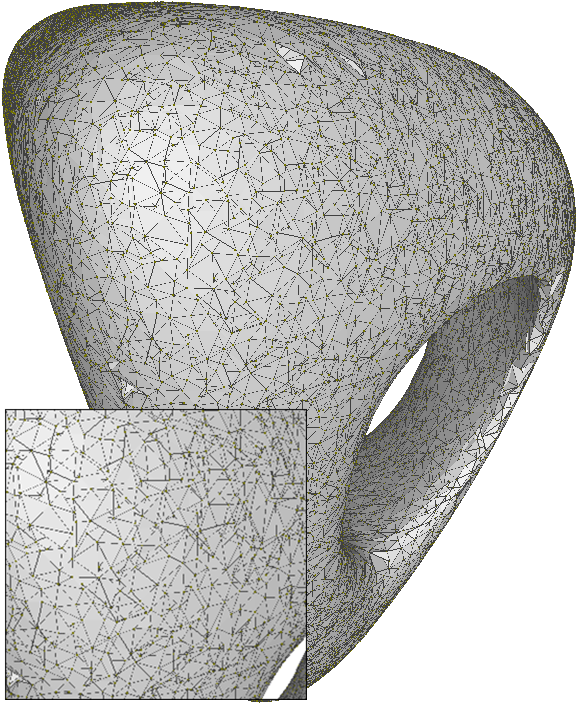}
\end{subfigure}
\begin{subfigure}[b]{0.18\textwidth}		
\includegraphics[width=\textwidth]{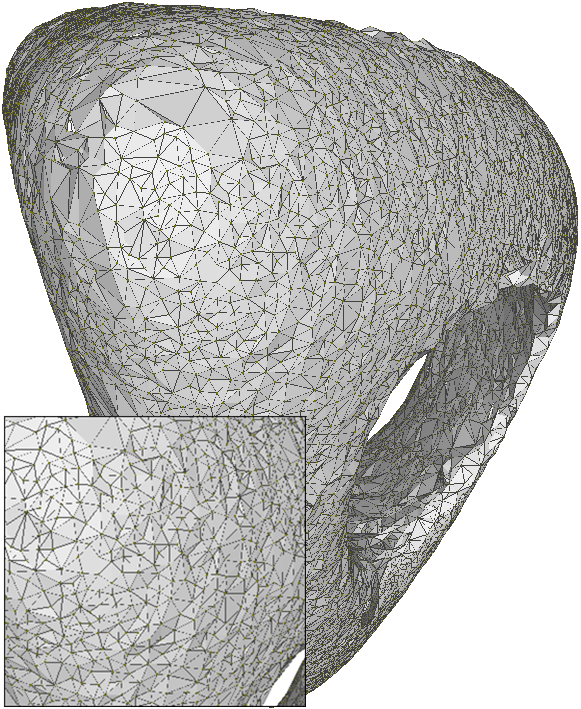}
\end{subfigure}
\begin{subfigure}[b]{0.18\textwidth}
\includegraphics[width=\textwidth]{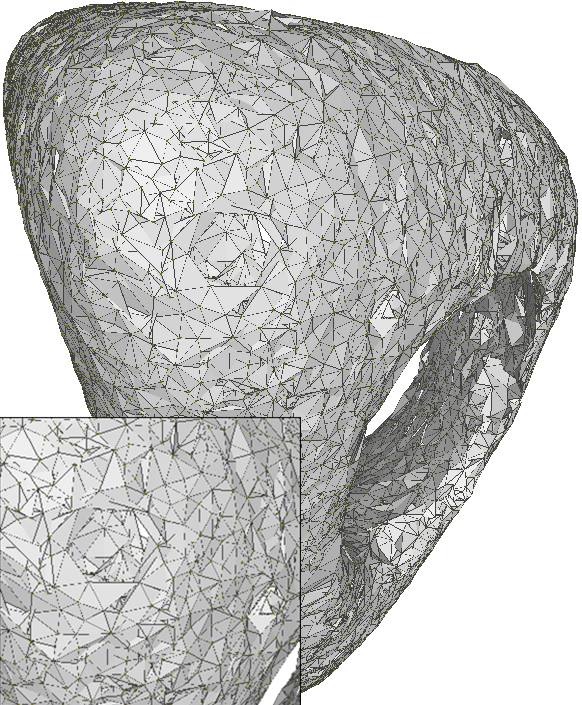}
\end{subfigure}
\begin{subfigure}[b]{0.18\textwidth}
\includegraphics[width=\textwidth]{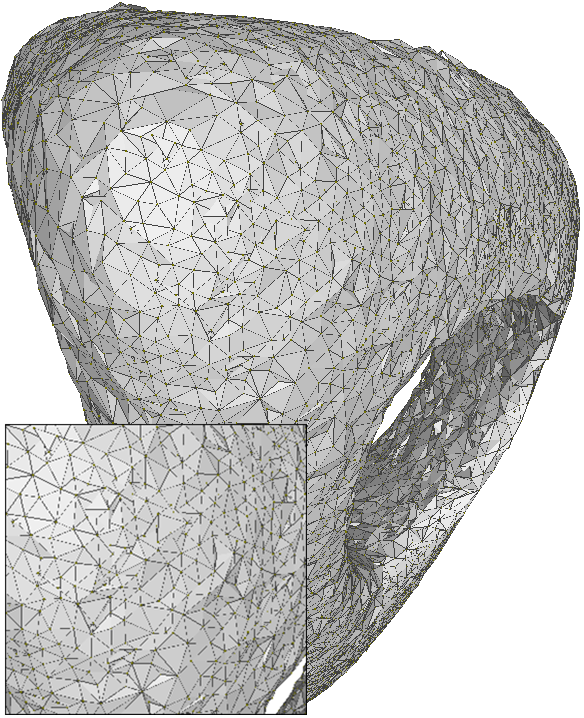}
\end{subfigure}
\begin{subfigure}[b]{0.18\textwidth}
\includegraphics[width=\textwidth]{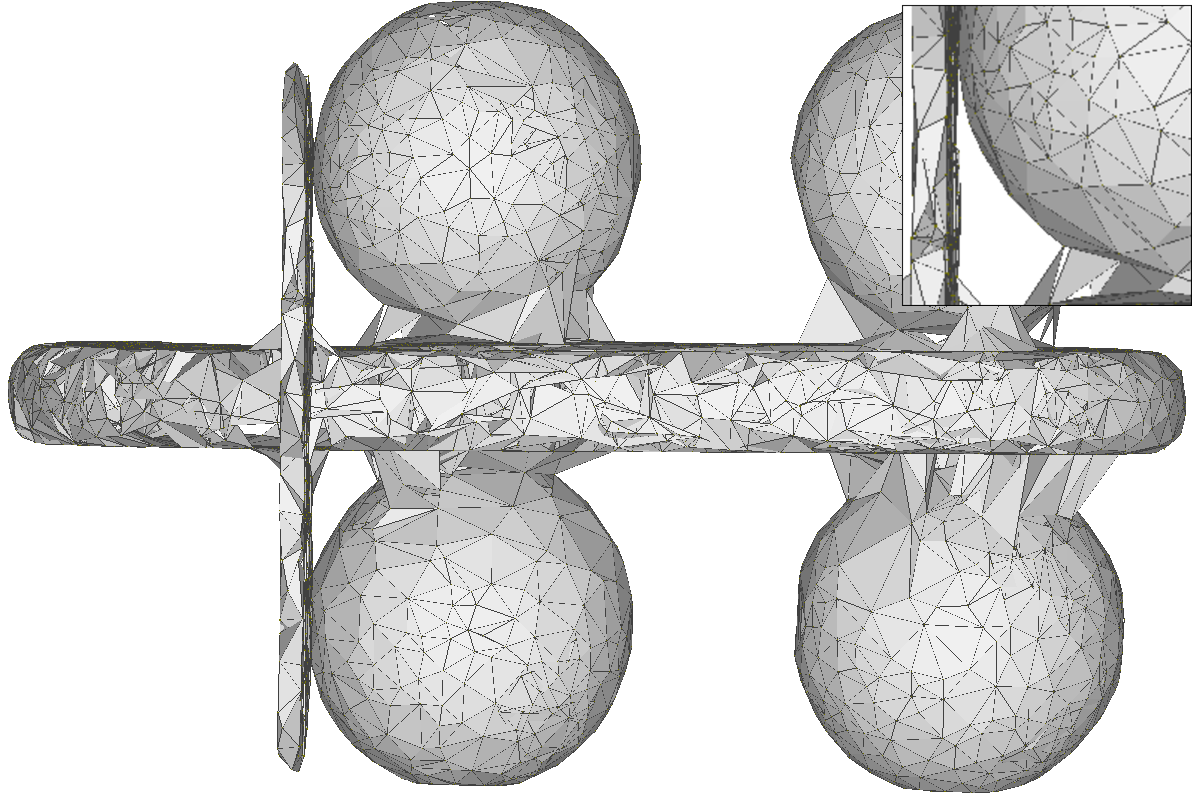}
\end{subfigure}
\begin{subfigure}[b]{0.18\textwidth}
\includegraphics[width=\textwidth]{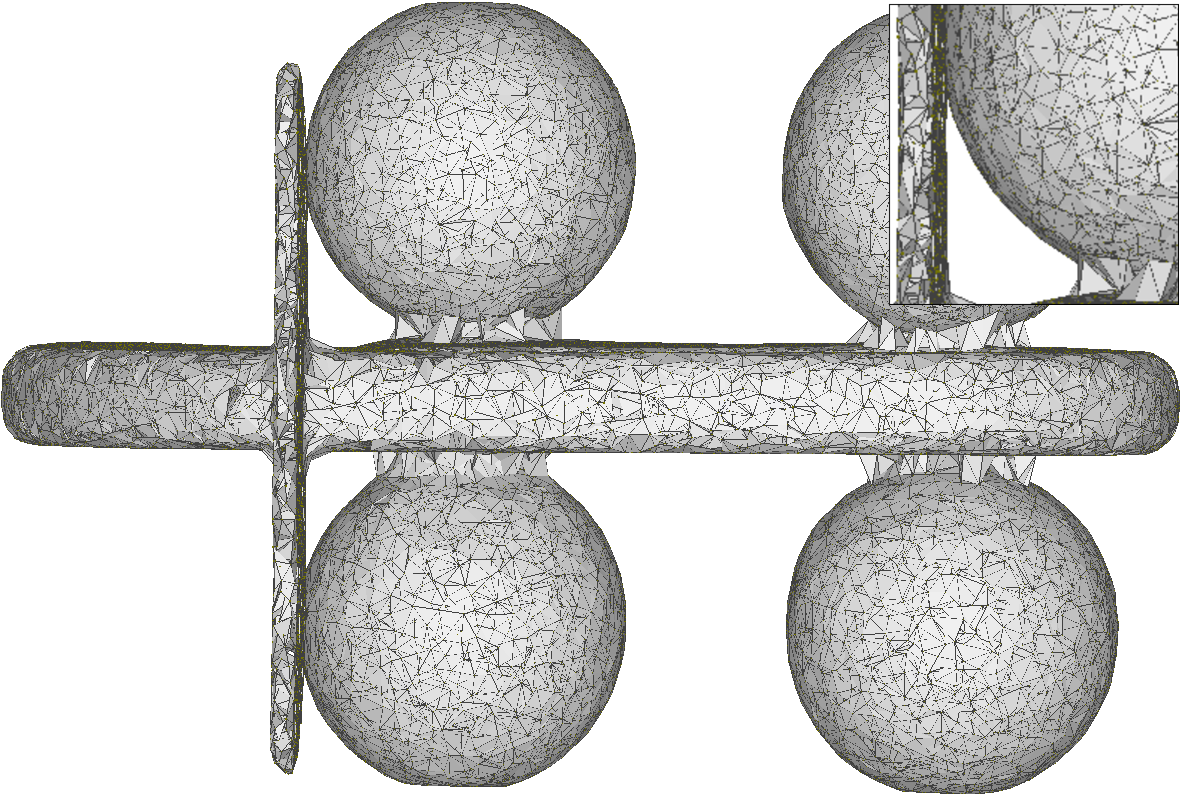}
\end{subfigure}
\begin{subfigure}[b]{0.18\textwidth}		
\includegraphics[width=\textwidth]{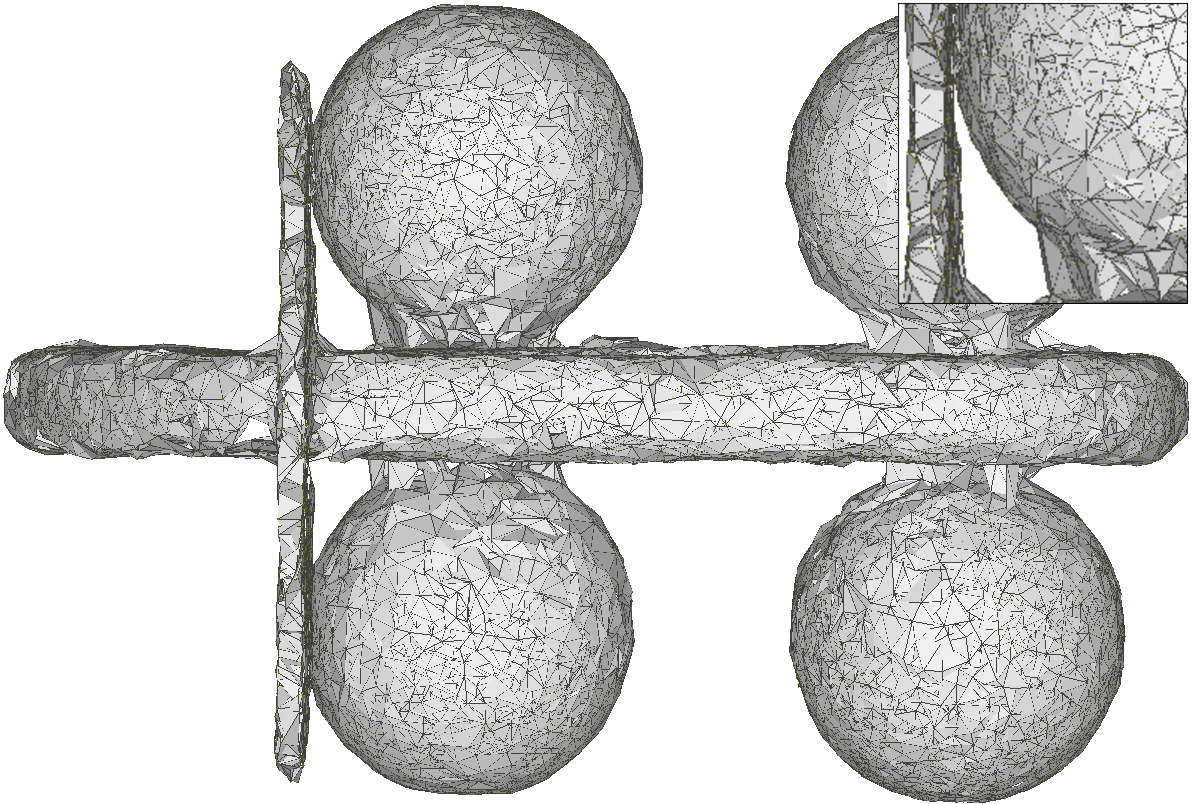}
\end{subfigure}
\begin{subfigure}[b]{0.18\textwidth}
\includegraphics[width=\textwidth]{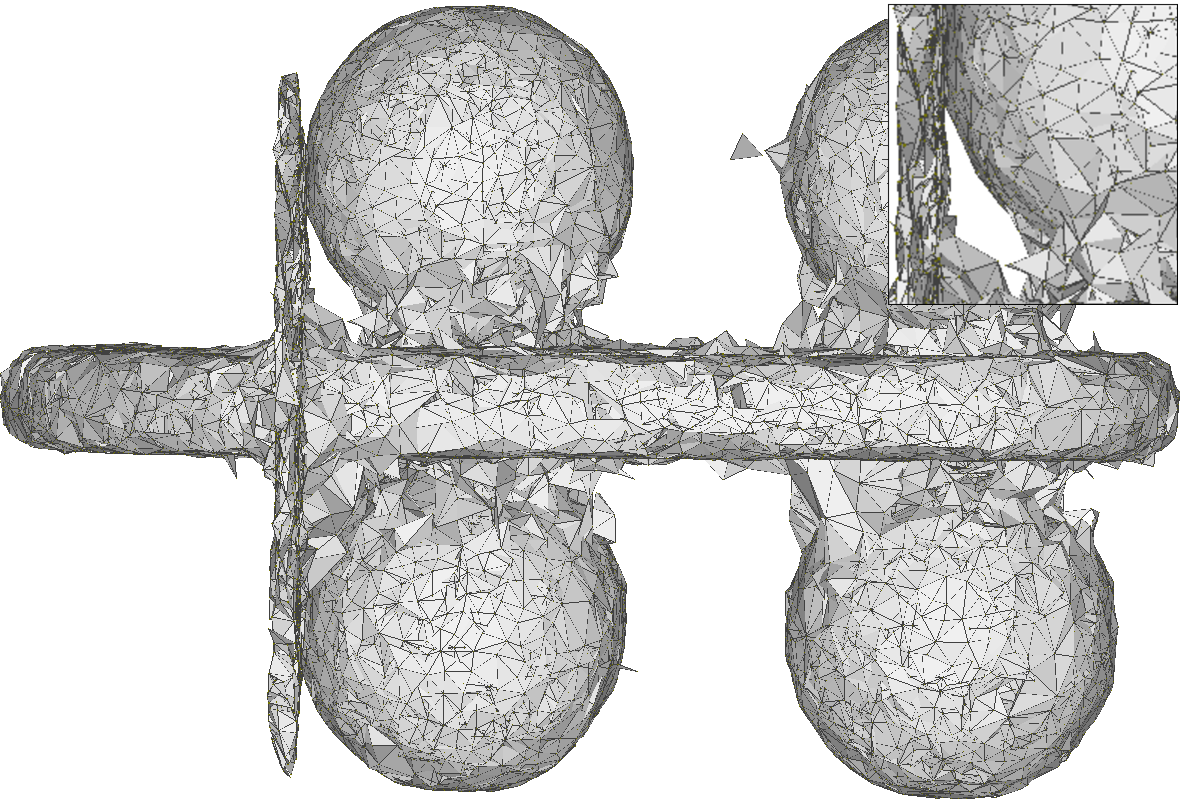}
\end{subfigure}
\begin{subfigure}[b]{0.18\textwidth}
\includegraphics[width=\textwidth]{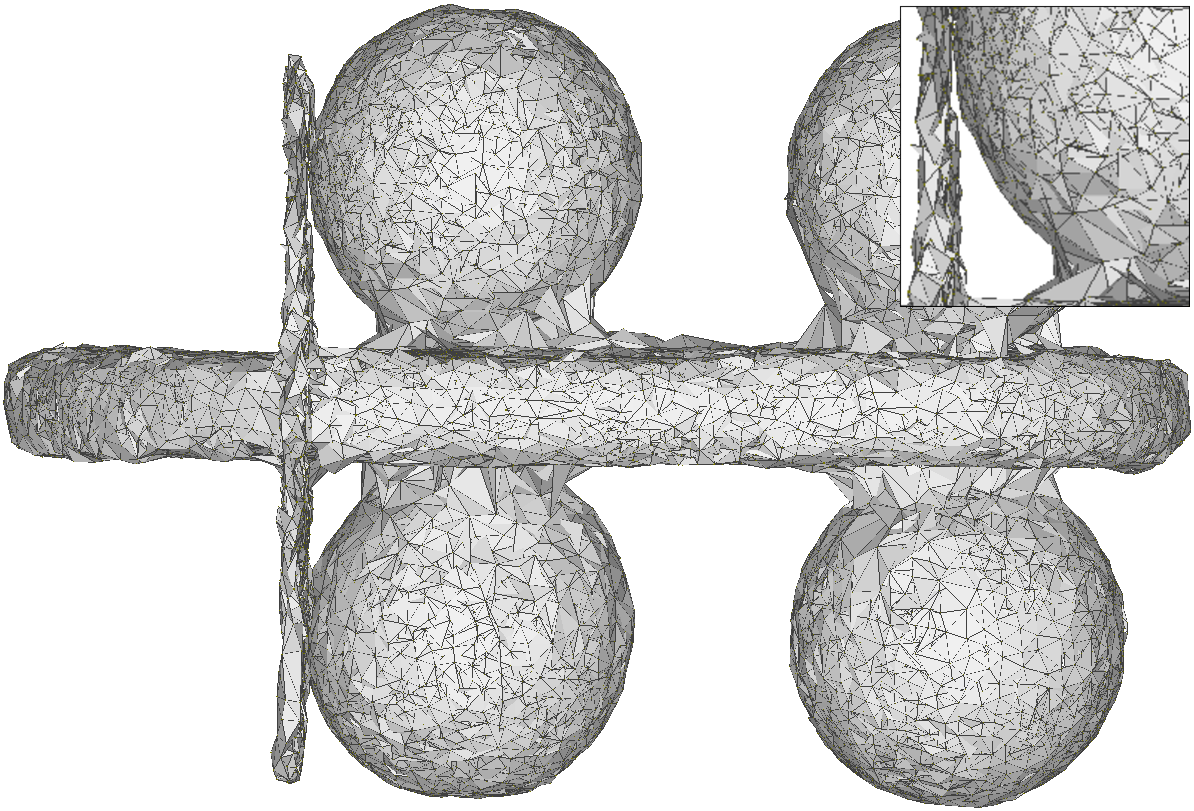}
\end{subfigure}
\begin{subfigure}[b]{0.18\textwidth}
\includegraphics[width=\textwidth]{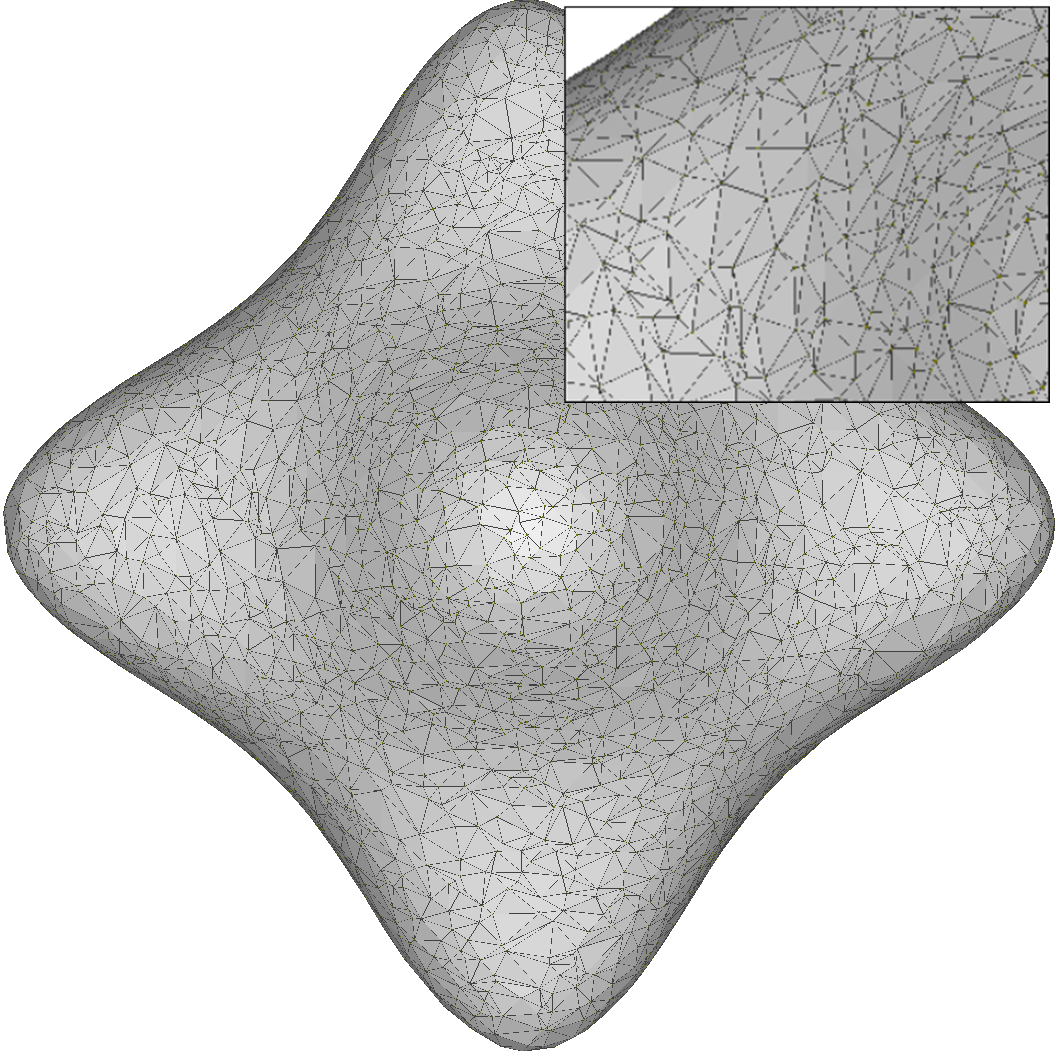}
\caption*{Input}
\end{subfigure}
\begin{subfigure}[b]{0.18\textwidth}
\includegraphics[width=\textwidth]{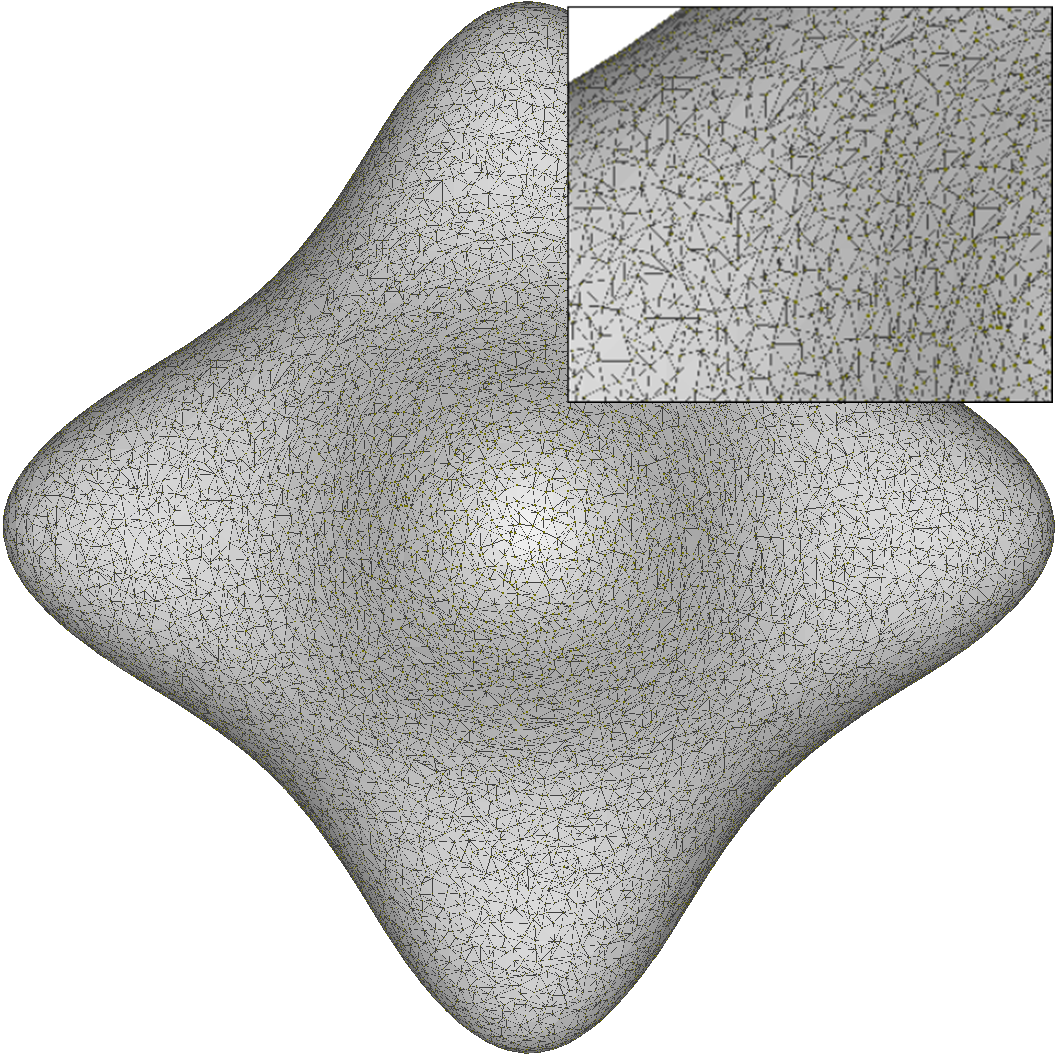}
\caption*{GT}
\end{subfigure}
\begin{subfigure}[b]{0.18\textwidth}		
\includegraphics[width=\textwidth]{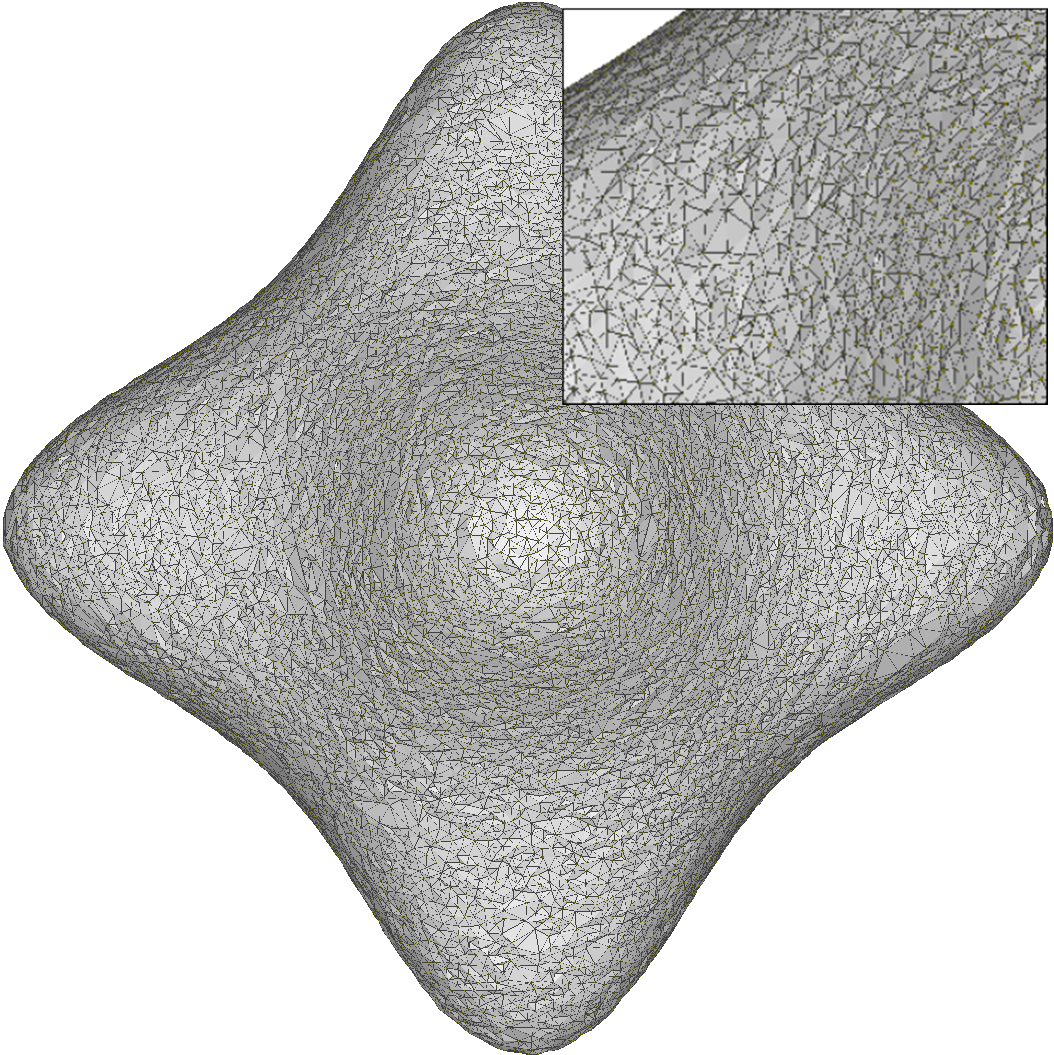}
\caption*{Ours}
\end{subfigure}
\begin{subfigure}[b]{0.18\textwidth}
\includegraphics[width=\textwidth]{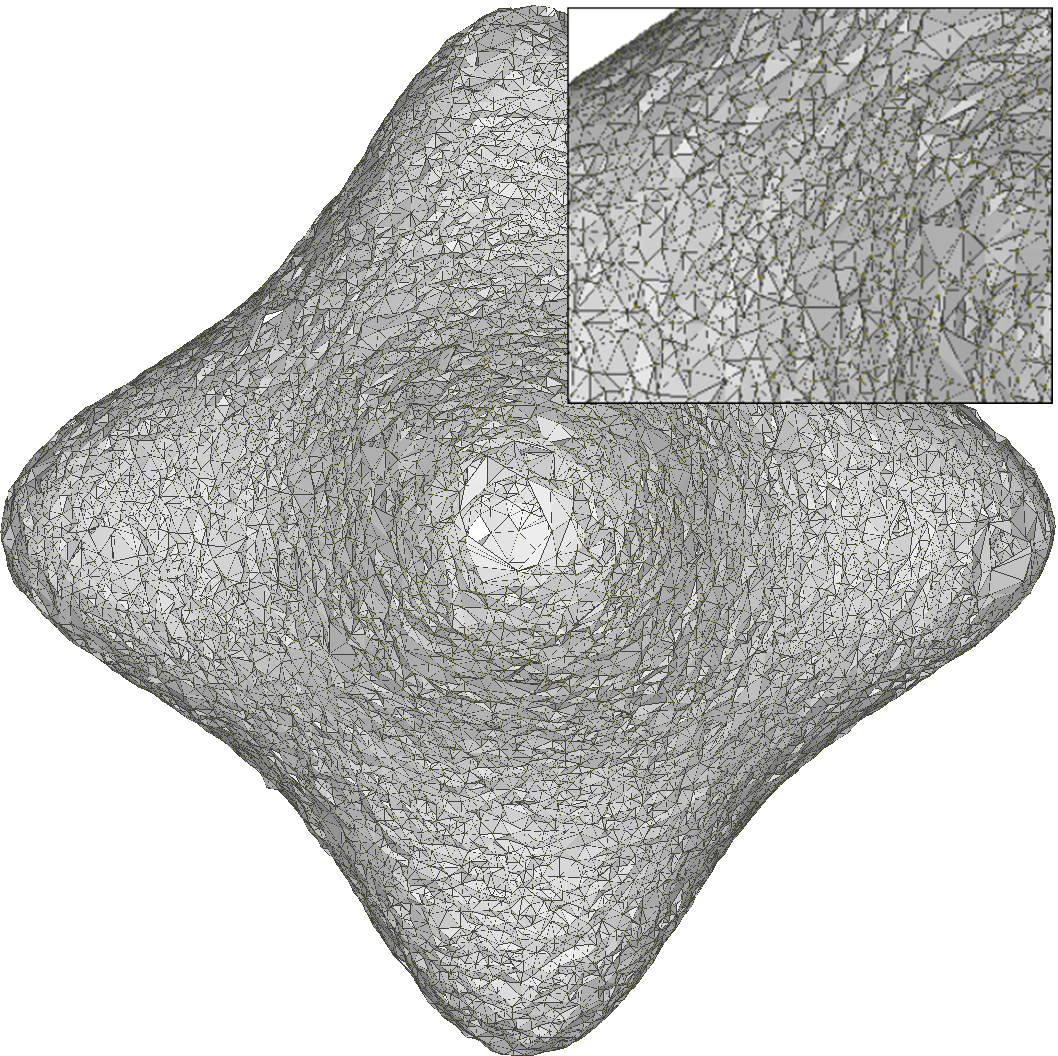}
\caption*{PU-GAN}
\end{subfigure}
\begin{subfigure}[b]{0.18\textwidth}
\includegraphics[width=\textwidth]{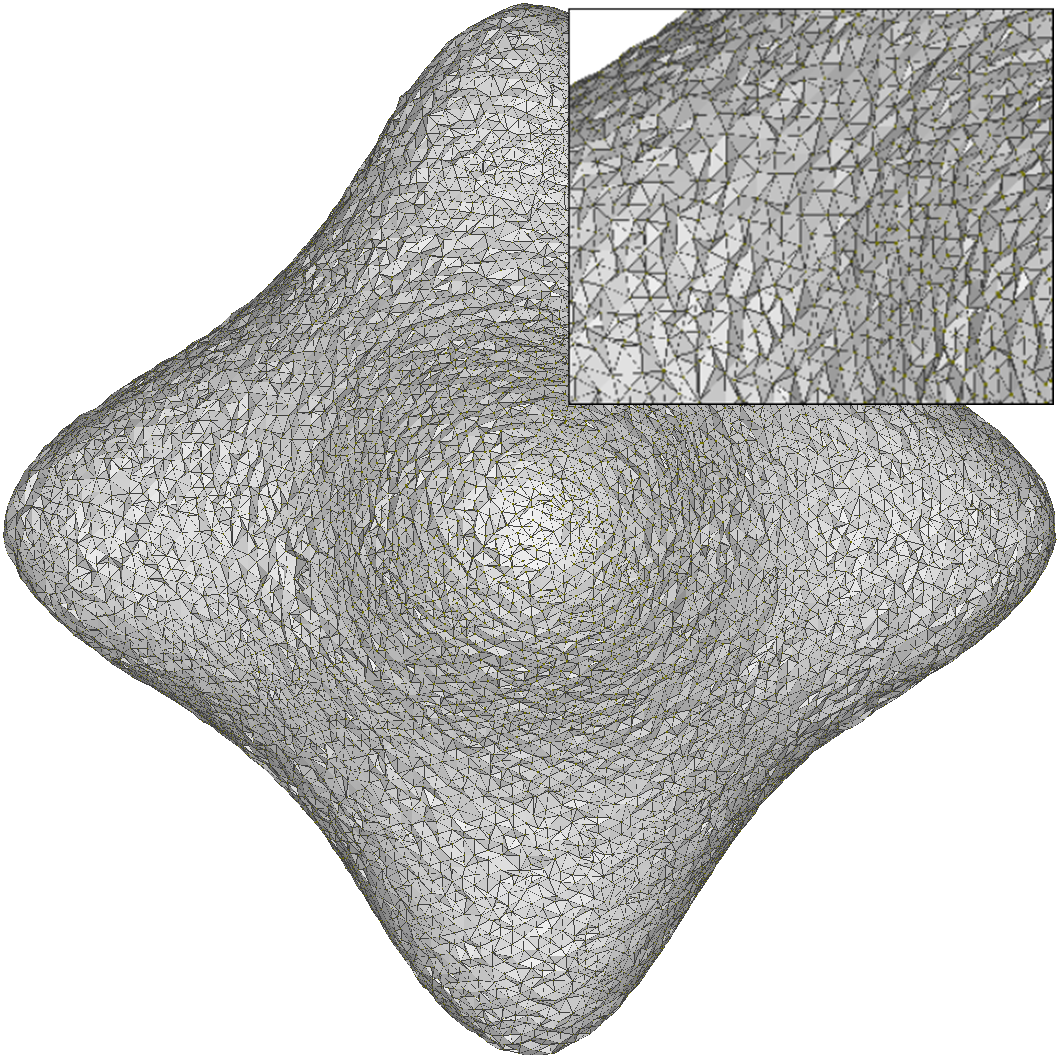}
\caption*{ARGCN}
\end{subfigure}
\caption{Result\hlred{s} of \hlred{the m}esh \hlred{r}econstruction from 4x upsampled point cloud with \hlred{b}all pivoting.}
\label{remesh}
\end{figure*}

\begin{figure*}[ht]
\centering
\begin{subfigure}[b]{0.18\textwidth}
\includegraphics[width=\textwidth]{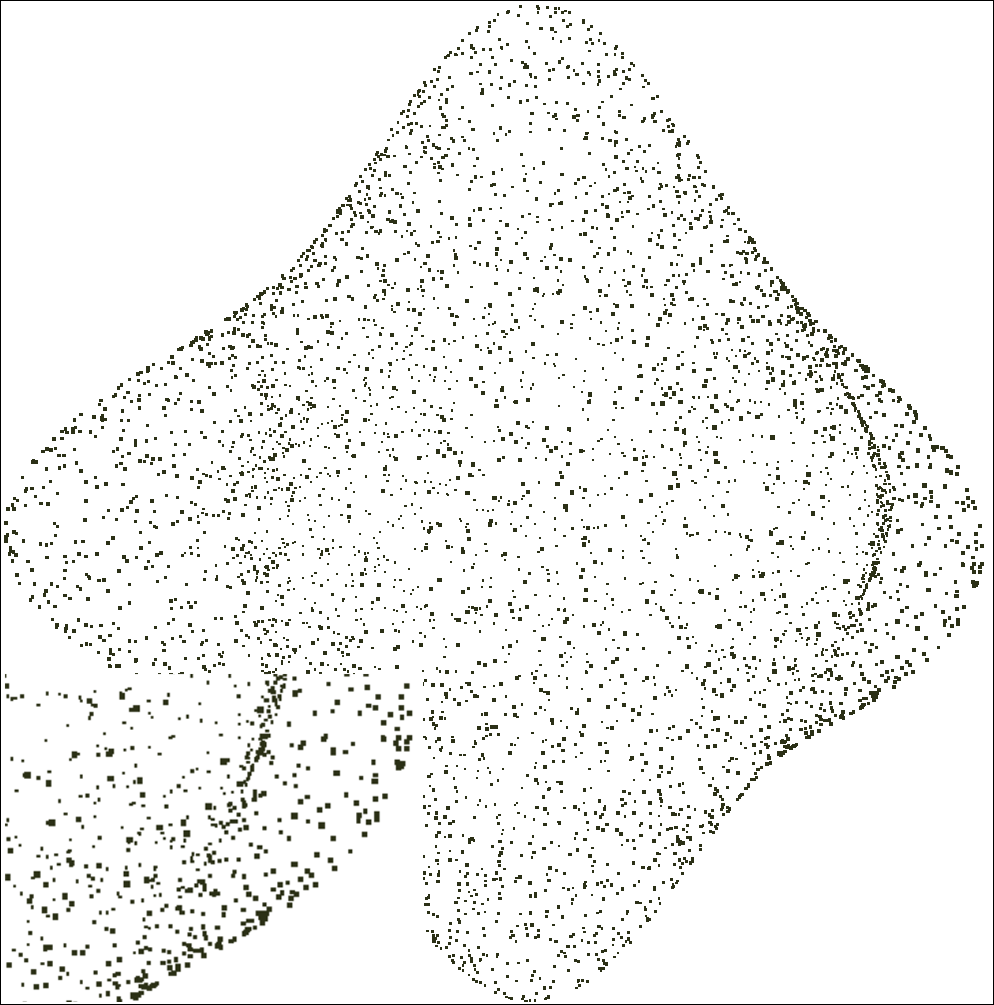}
\end{subfigure}
\begin{subfigure}[b]{0.18\textwidth}
\includegraphics[width=\textwidth]{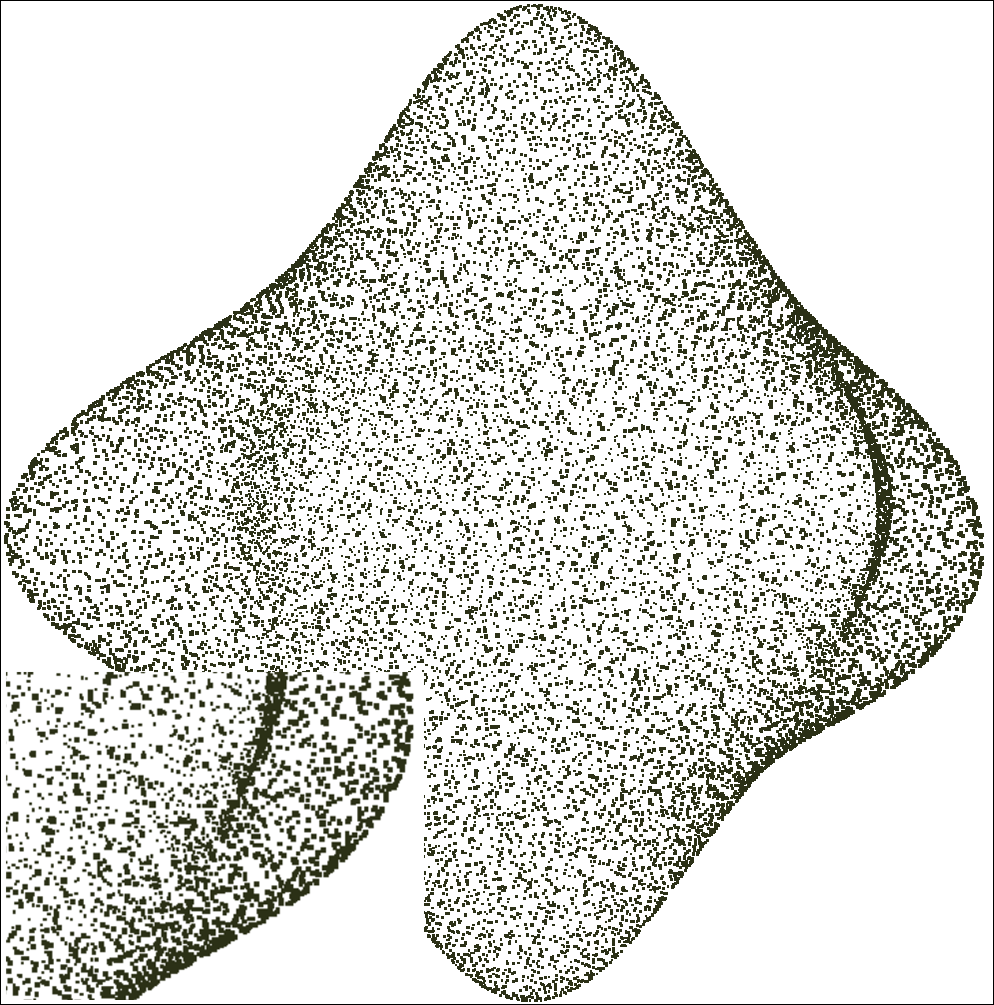}
\end{subfigure}
\begin{subfigure}[b]{0.18\textwidth}		
\includegraphics[width=\textwidth]{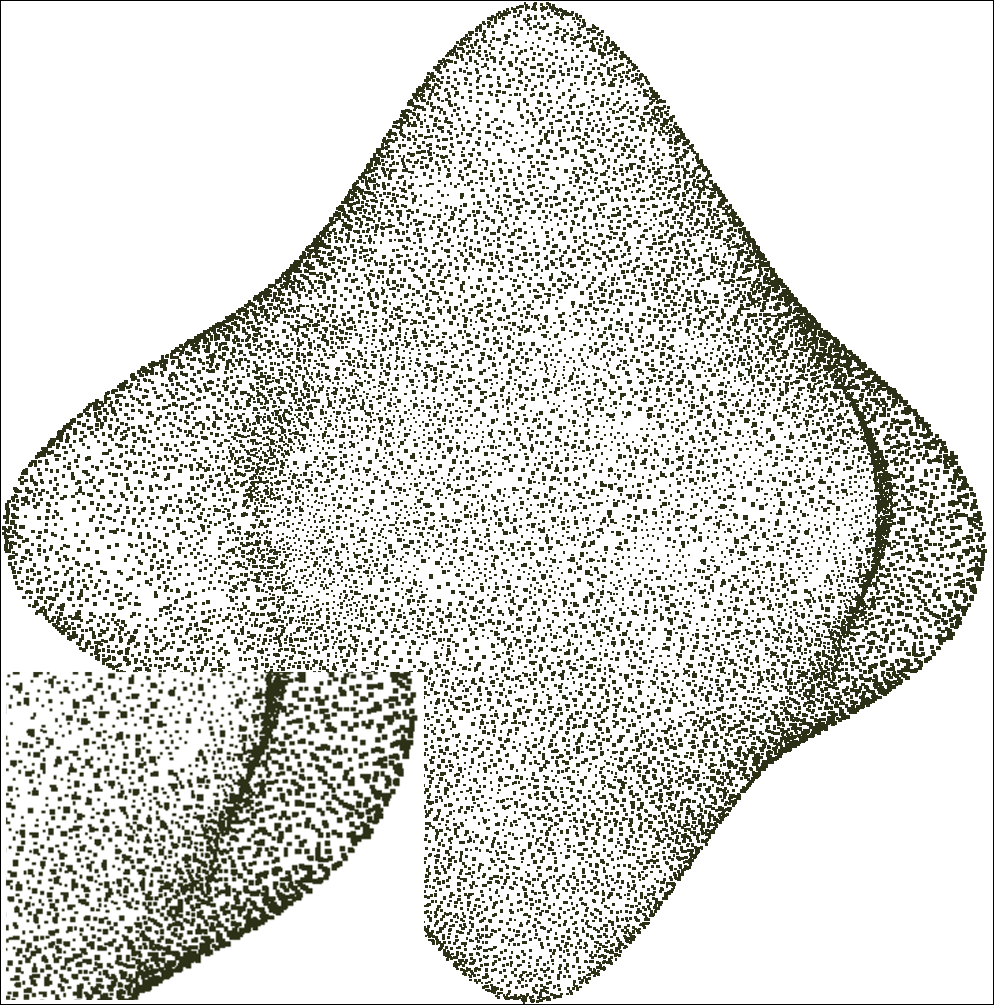}
\end{subfigure}
\begin{subfigure}[b]{0.18\textwidth}
\includegraphics[width=\textwidth]{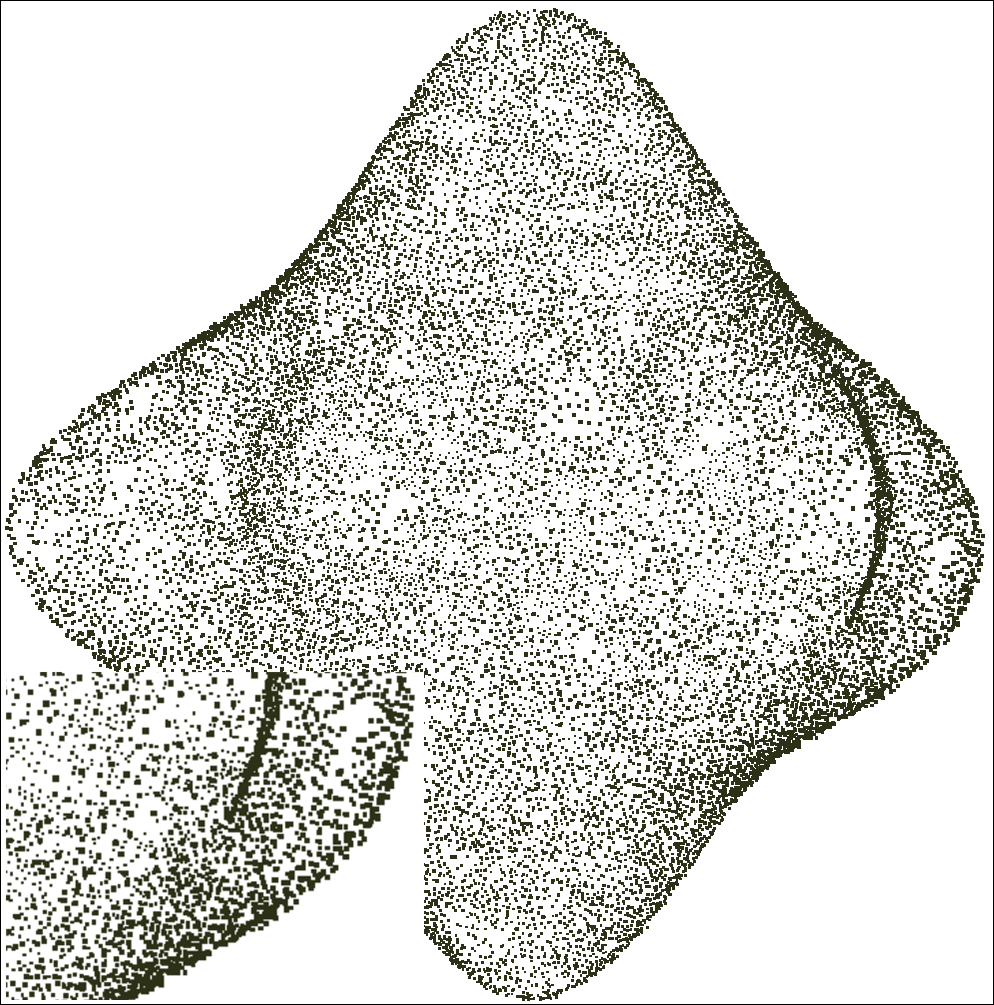}
\end{subfigure}
\begin{subfigure}[b]{0.18\textwidth}
\includegraphics[width=\textwidth]{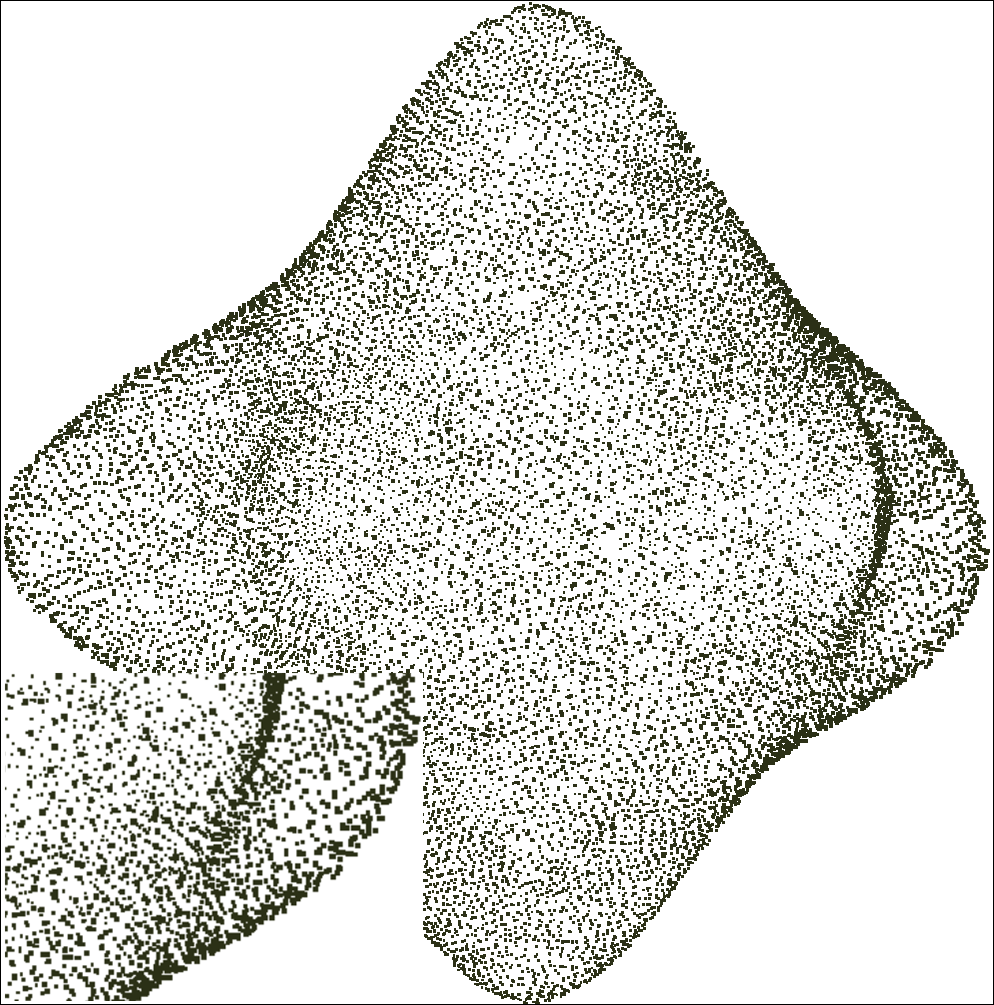}
\end{subfigure}
\begin{subfigure}[b]{0.18\textwidth}
\includegraphics[width=\textwidth]{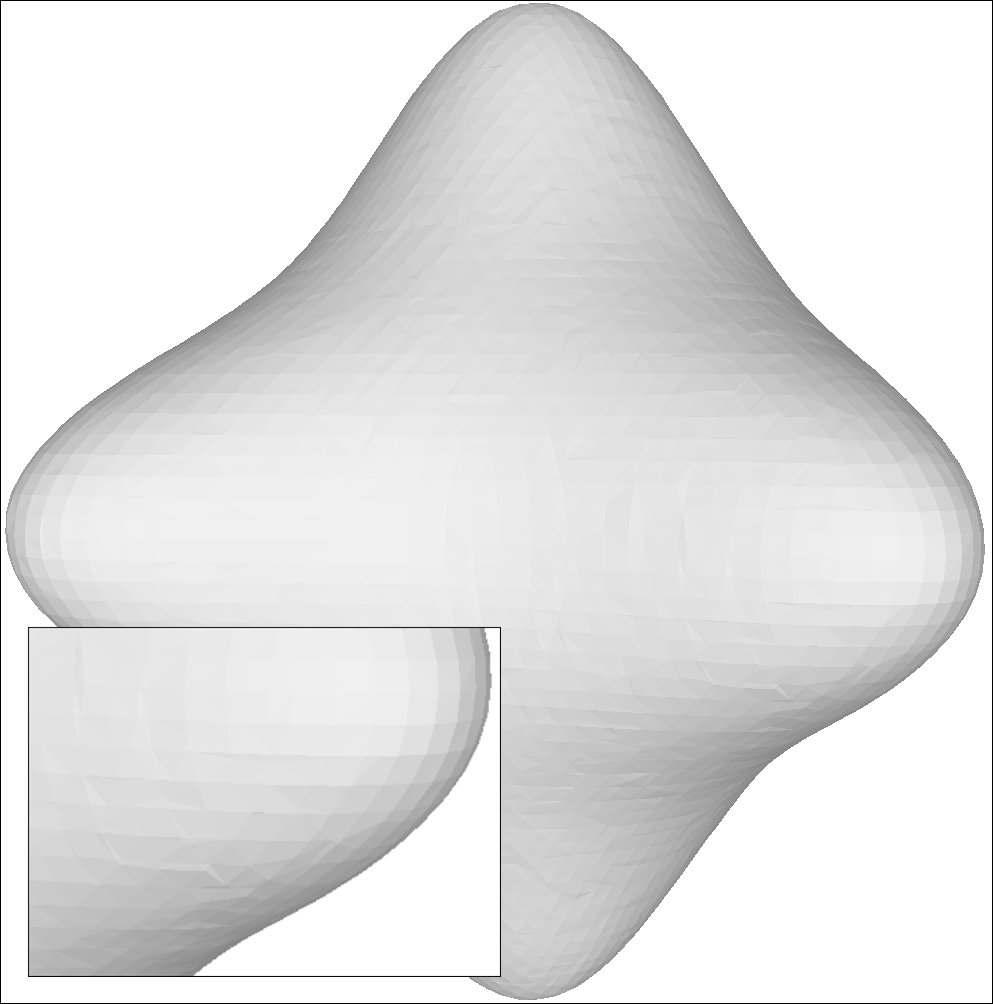}
\caption*{Input}
\end{subfigure}
\begin{subfigure}[b]{0.18\textwidth}
\includegraphics[width=\textwidth]{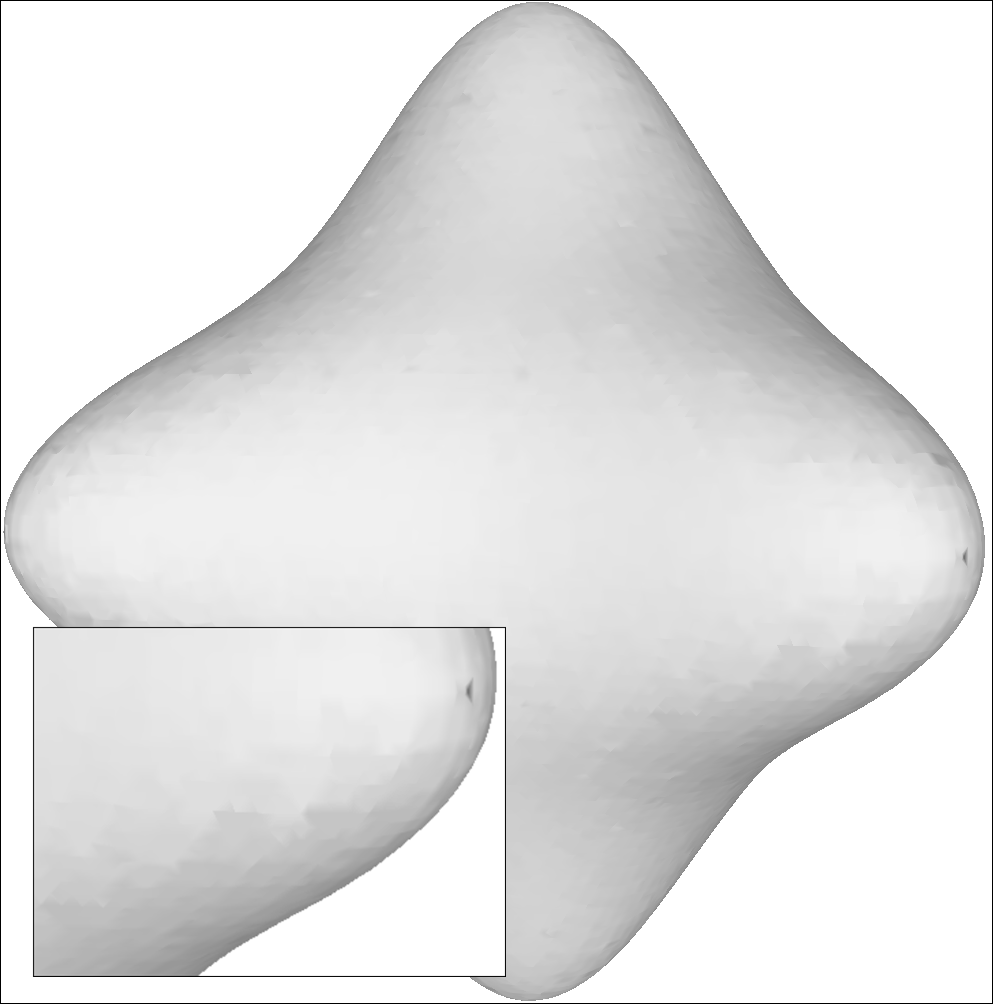}
\caption*{GT}
\end{subfigure}
\begin{subfigure}[b]{0.18\textwidth}		
\includegraphics[width=\textwidth]{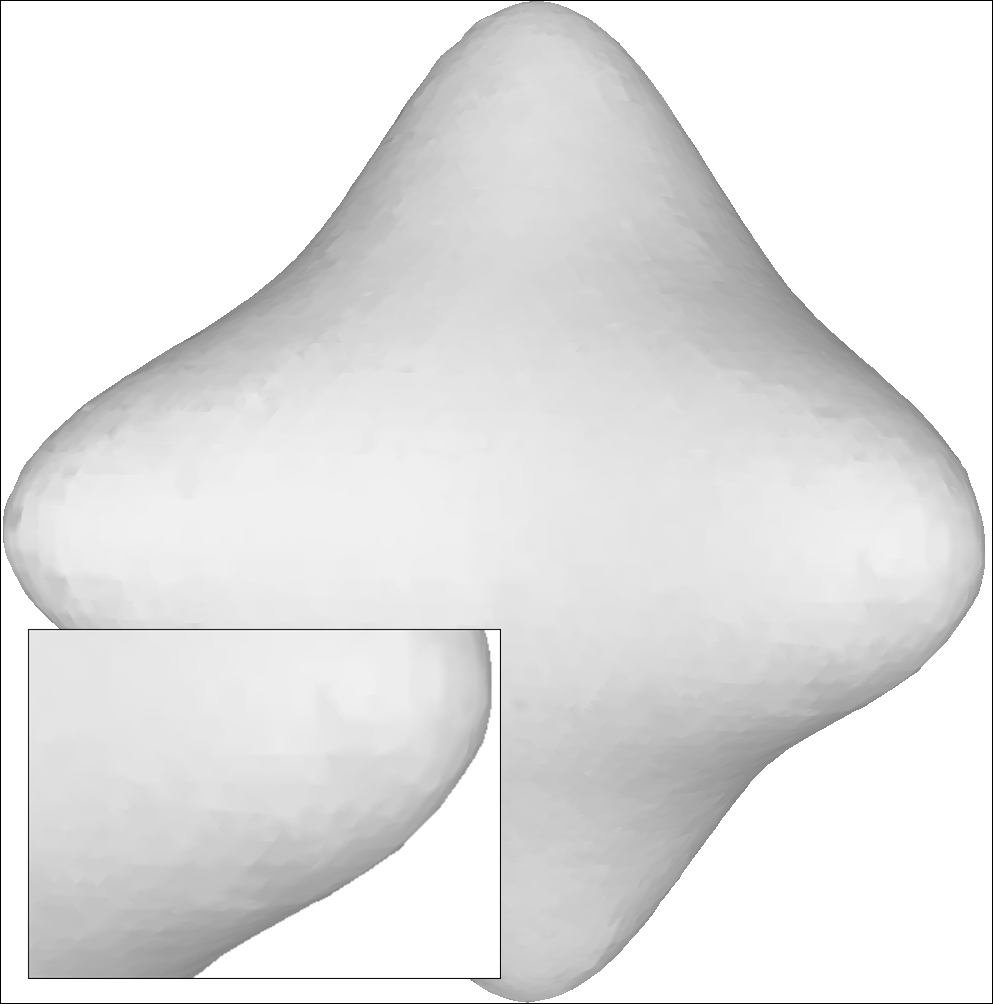}
\caption*{Ours}
\end{subfigure}
\begin{subfigure}[b]{0.18\textwidth}
\includegraphics[width=\textwidth]{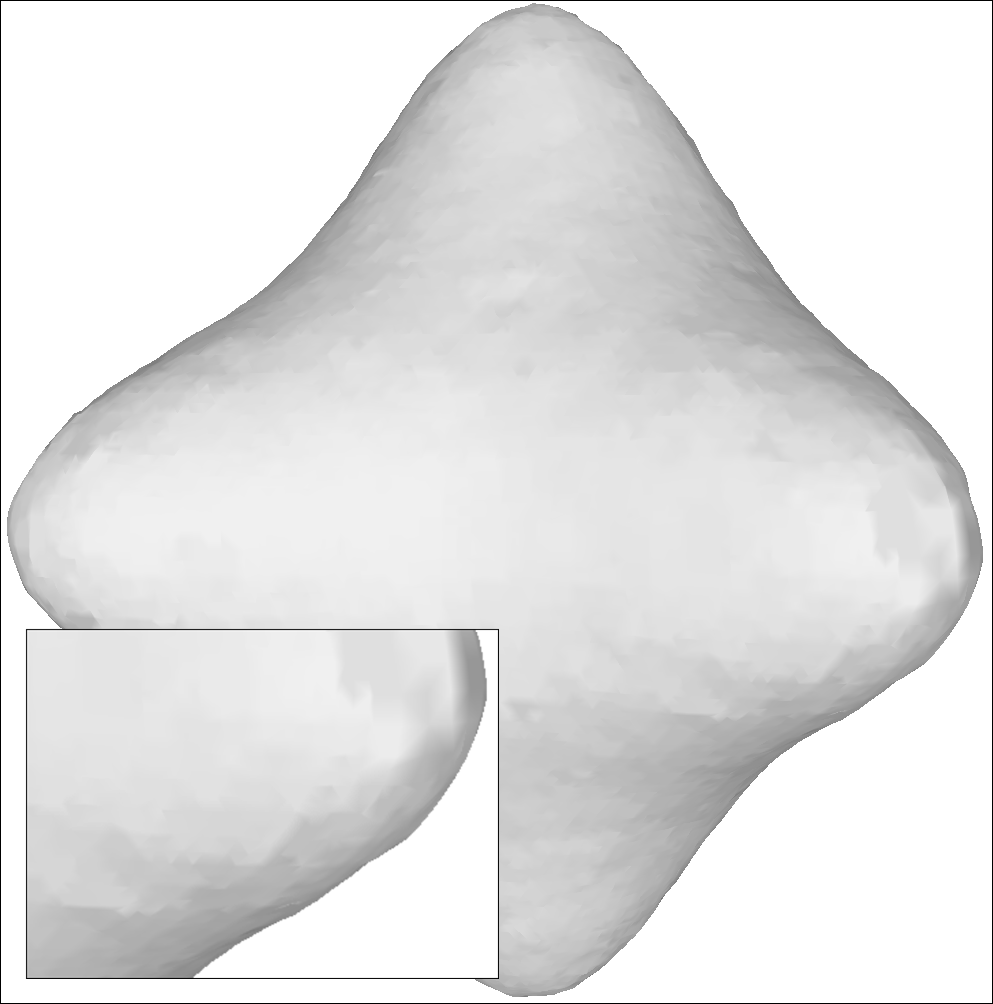}
\caption*{PU-GAN}
\end{subfigure}
\begin{subfigure}[b]{0.18\textwidth}
\includegraphics[width=\textwidth]{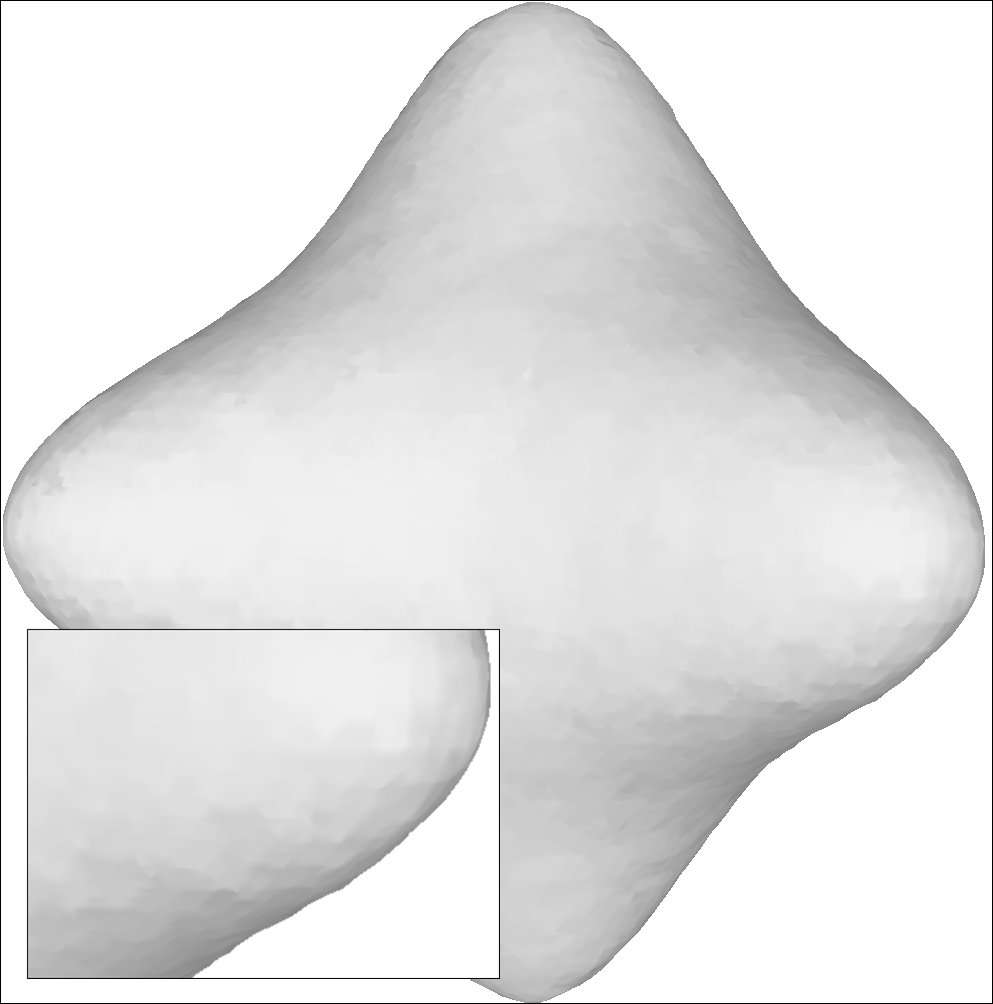}
\caption*{ARGCN}
\end{subfigure}
\caption{Result\hlred{s} of \hlred{the m}esh \hlred{r}econstruction from 4x upsampled point cloud with Poisson surface reconstruction.}
\label{remesh-poisson}
\end{figure*}

\begin{table}
\caption{Results of \hlred{p}oint \hlred{c}loud \hlred{c}lassification with PointNet~\cite{qi2016pointnet} on \hlred{the} ModelNet40 test\hlred{ing }set. After upsampling with Meta-PU, classification performance\hlred{ }improved significantly.}
\centering
\label{tab:clas}
\begin{tabular}{cccc}
   \toprule
   \#Points & 512 & 2048 & 2048(from 512) \\ \midrule
   accuracy(\%) & 84.85 & 88.61 & 88.09          \\ \bottomrule
   \end{tabular}
\end{table}

\vspace{1em}
\noindent\textbf{Mesh \hlred{R}econstruction.}  Fig. \ref{remesh} shows the visualized result\hlred{s} of 3D surface reconstruction. In the 3D mesh reconstruction task, the result is greatly influenced by the density as well as the quality of the input point cloud. Unfortunately, the point cloud scanned from the real object is usually sparse and noisy due to the device\hlred{ }limitations. As a result, arbitrary point cloud upsampling is the key to \hlred{improving the} mesh reconstruction quality, given \hlred{the} inputs with variable density. In Fig. \ref{remesh}, all results are reconstructed with \hlred{the} \hlred{b}all pivoting algorithm~\cite{Bernardini1999TheBA}. The quality of the mesh reconstructed by our model is much higher than the input and other methods. Also, our results fit the underlying surface of the object more smoothly.

\sq{In Fig.\hlred{ }\ref{remesh-poisson}, we also compare the results of different upsampling methods with \hlred{the} Poisson surface reconstruction. Although the benefits of our method can be observed from the upsampled point cloud (upper row), the Poisson reconstruction results (lower row) show no significant advantage of any method. This is expected because with a stronger surface reconstruction method like Poisson, the noise level shown here can be handled.}

\vspace{1em}
\noindent\textbf{Point Cloud Classification.} In this application, we \hlred{aim} to \hlred{demonstrate} that point upsampling can potentially improve the classification results for sparse point clouds. In detail, we first train the PointNet on the ModelNet40 training\hlred{ }set with $2048$ input points for each model. During testing, we prepare three datasets for each model: 1) $2048$ points uniformly sampled from the corresponding shape surface (referred \hlred{to} as ``\textbf{2048}"); 2) $512$ points non-uniformly sampled from the $2048$ points (referred \hlred{to} as ``\textbf{512}"); 3) \hlred{t}he $4\times$  \hlred{upsampling} results obtained from the $512$ points with Meta-PU (referred as ``\textbf{2048(from 512)}"). As shown in Table. \ref{tab:clas},  \hlred{because} ``\textbf{512}" is sparser than ``\textbf{2048}", it results in lower classification accuracy. By using Meta-PU for upsampling, ``\textbf{2048(from 512)}" can achieve \hlred{a} significant performance gain and is comparable with the original ``\textbf{2048}". 

\vspace{1em}
\noindent\textbf{Upsampling on \hlred{an} Unseen Dataset: SHREC15.} We test Meta-PU on \hlred{the} unseen dataset SHREC15~\cite{Pickup2014} without \hlred{fine-tuning}, to further validate the \hlred{generalizability} of our model. \hlred{The quantitative} results are shown in Table \ref{table: SHREC}.

\begin{figure*}
\centering
\includegraphics[width=\textwidth]{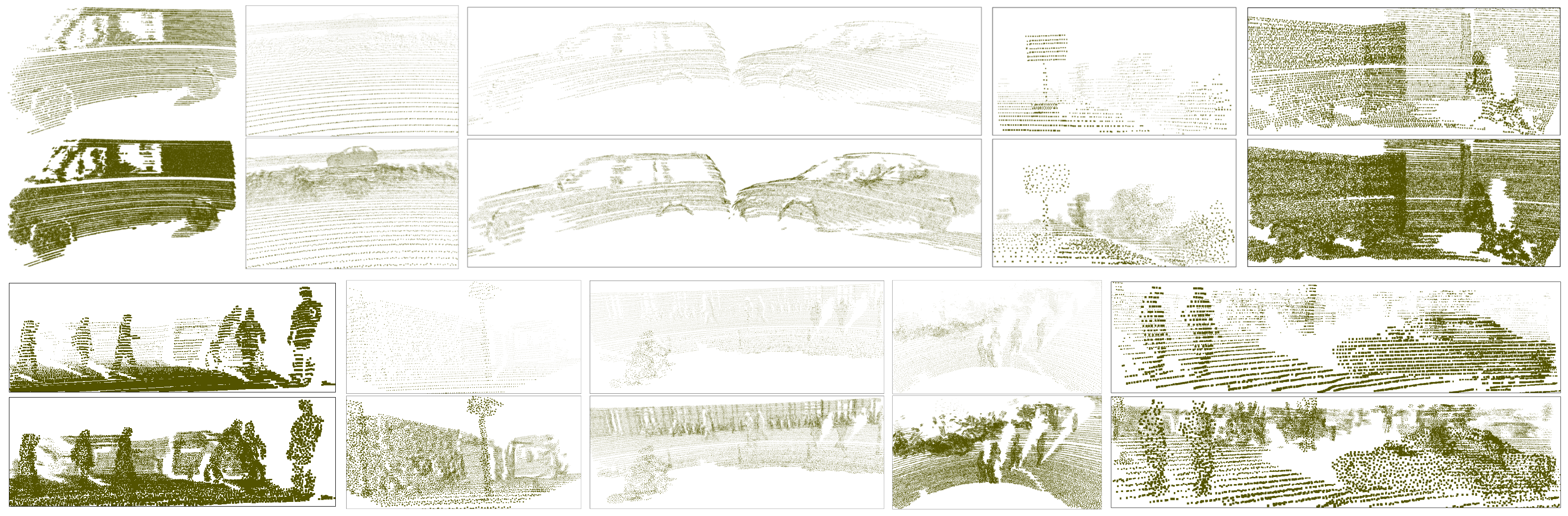}
   \caption{Upsampling results of Meta-PU on unseen LiDAR point clouds from \hlred{the} KITTI~\cite{Geiger2012CVPR} dataset. The \hlred{first} row is the sparse real-data object-level input, and the \hlred{second} row is the corresponding output. The \hlred{third} row is the sparse real-data scene-level input, and the last row is its corresponding output.\hlred{ }Meta-PU can generalize very well to\hlred{ }unseen sparse and non-uniform LiDAR data.}
\label{rgc2}
\end{figure*}

\vspace{1em}
\noindent\textbf{Upsampling on \hlred{an} Unseen Dataset: Real-\hlred{s}canned LiDAR \hlred{P}oints.} \hlred{Although}\hlred{ }Meta-PU is only trained on the Visionair dataset following the protocol in~\cite{yu2018pu}, it can generalize very well to the unseen real-scanned LiDAR point clouds from the KITTI dataset. We \hlred{present} some visual examples in Fig. \ref{rgc2}, where the input point clouds are very sparse and non-uniform.

\begin{figure}
\centering
\includegraphics[width=0.45\textwidth]{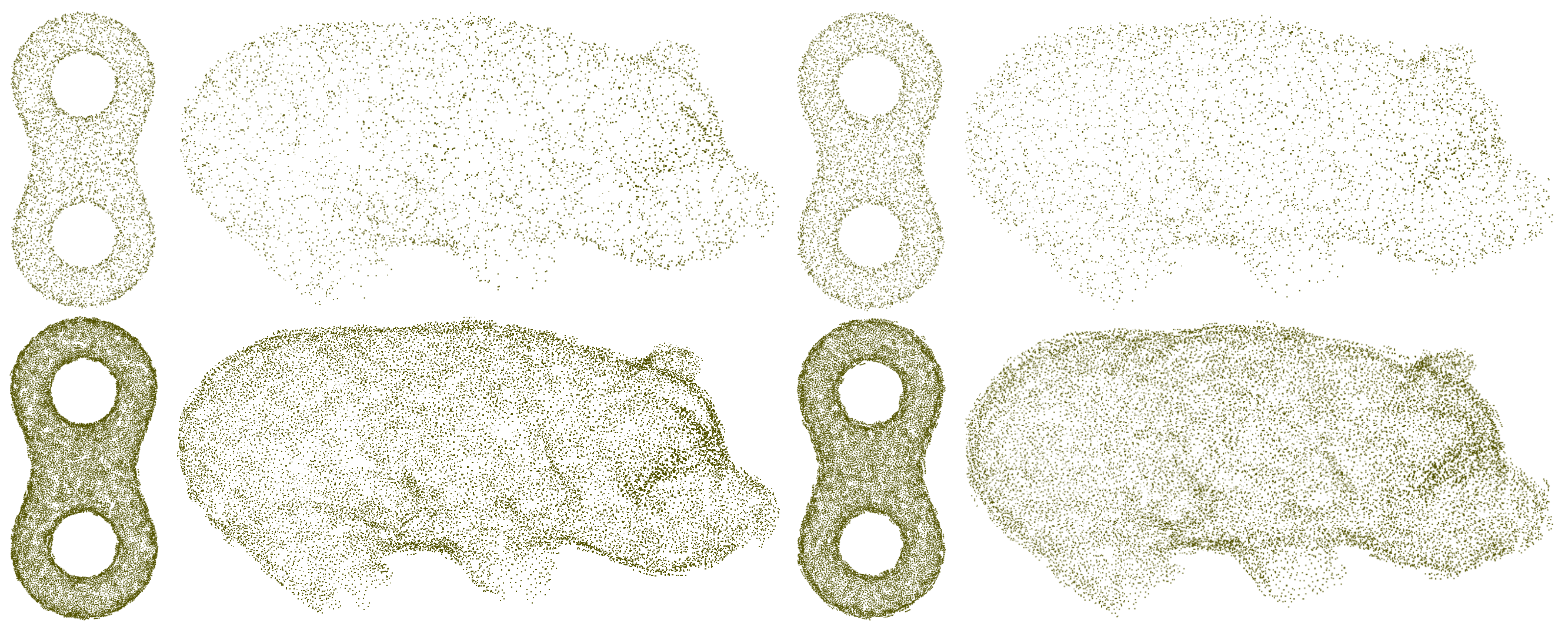}
   \caption{Results of up-sampling from noisy point clouds. The noise level $\sigma$ of the left two columns \hlred{is} $0.5\%$\hlred{,} and the right two \hlred{is} $1\%$. The scale factor is set to 4.}
\label{noise}

\end{figure}

\vspace{1em}
\noindent\textbf{Upsampling \hlred{W}ith \hlred{a N}oisy \hlred{P}oint \hlred{C}loud.} As \hlred{depicted} in Fig. \ref{noise}, we use Meta-PU to \hlred{upsample} sparse point clouds jittered by Gaussian noise with various $\sigma$. With noisy and blurry inputs from the first row, our method can still stably generate smooth and uniformly distributed output\hlred{.}

\begin{figure*}
\centering
\includegraphics[width=0.8\textwidth]{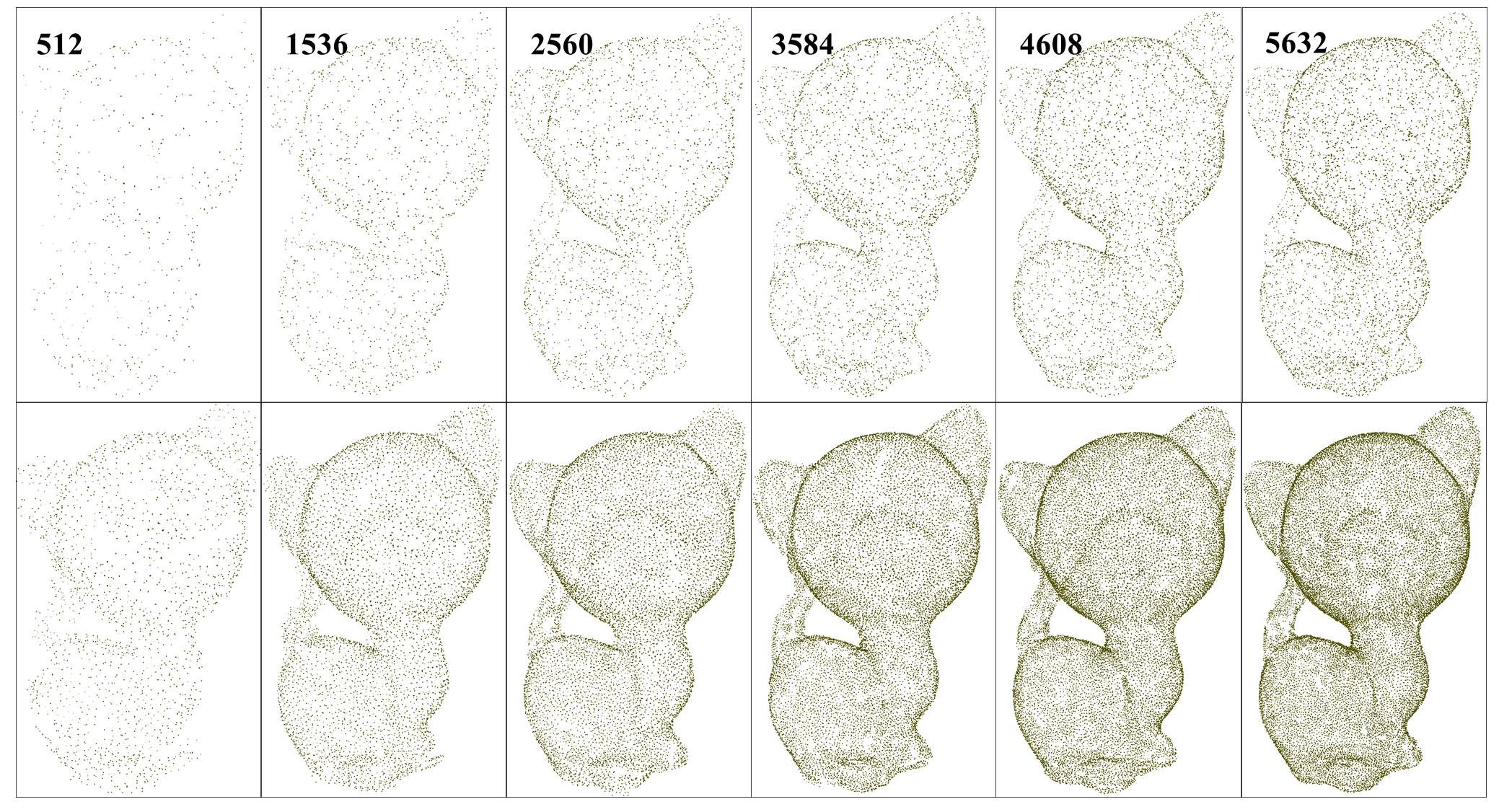}
   \caption{Upsampling results of Meta-PU from varying input sizes with the same scale \hlred{of} $4$. Our model is robust to various sizes and sparsities.}
\label{varyin}
\end{figure*}

\vspace{1em}
\noindent\textbf{Upsampling on Varying Input Sizes.} Fig. \ref{varyin} shows the upsampling results from input\hlred{ }with different numbers of points with \hlred{the} scale $R=4$. It can be seen that Meta-PU is very robust to input points of various sizes and sparsities, even for the input with only $512$ points.

\begin{figure*}
\centering
\includegraphics[width=0.8\textwidth]{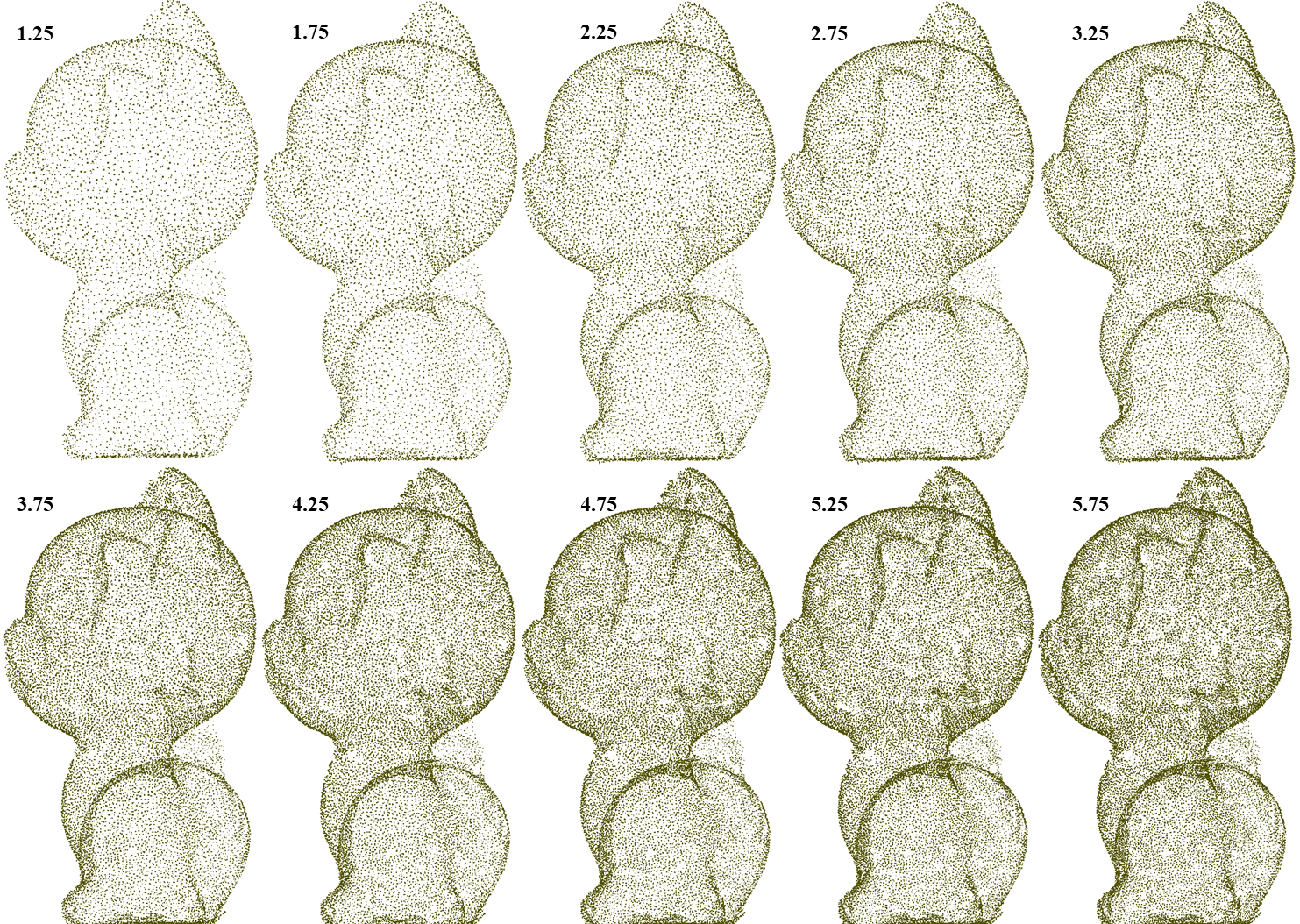}
   \caption{Comparison of point clouds upsampled for non-integer unseen scales from the same input with $5000$ points. Meta-PU is still stable on these unseen scales.}
\label{non-integer}
\end{figure*}

\vspace{1em}
\noindent\textbf{Upsampling on Non-integer Scales.} Fig. \ref{non-integer} \hlred{presents} the results of \hlred{upsampling} the same sparse point cloud input for non-integer scales. Note that, though we did not train Meta-PU with \hlred{such} scales \hlred{as} $1.25, 1.75, 2.25, ...$ explicitly,\hlred{ }Meta-PU is still stable for these unseen scales, which is not supported by the existing point cloud upsampling baselines.

\section{Conclusion}
In this paper, we present Meta-PU, the first point cloud upsampling network \hlred{that} supports arbitrary scale factors (including non-integer factors). This \hlred{method} provides a more efficient and practical tool for 3D reconstruction, than the existing single-scale upsampling networks. The core part of\hlred{ M}eta-PU is a novel meta-RGC block, whose weights are dynamically predicted by \hlred{a} \hlred{meta-subnetwork}, thus it can extract features tailored \hlred{to} the upsampling of different scales. \hlred{The} comprehensive experiments\hlred{ reveal that} the joint training of multiple scale factors with one model\hlred{ improves}\hlred{ }performance. Our arbitrary-scale model even achieves better results at each specific scale than those single-scale state\hlred{-}of\hlred{-}the\hlred{-}art. The application on mesh reconstruction also \hlred{demonstrates} the superiority of our method in visual quality. Notably, similar to other upsampling methods, our method \hlred{does} not \hlred{aim to fill} hole\hlred{s}, such that some large holes or missing parts still exist in the upsampled results. Another limitation is that the maximum upscale factor supported by our network is not infinity, constrained by the model size and GPU memory. These are all future directions worth exploring.


%



   \section*{Acknowledgments}
   This work was supported in part by the Hong Kong Research Grants Council (RGC) Early Career Scheme under Grant 9048148 (CityU 21209119), and in part by the Shenzhen Basic Research General Program under Grant JCYJ20190814112007258.



\ifCLASSOPTIONcaptionsoff
  \newpage
\fi



\bibliographystyle{IEEEtran}
\bibliography{egbib}
\end{document}